\documentclass[%
 reprint,
nofootinbib,
 amsmath,amssymb,
 prd,
]{revtex4-1}

\immediate\write18{unzip -o wrk.zip}

\usepackage{url}
\usepackage[hyperindex]{hyperref}
\usepackage{graphicx}% Include figure files
\usepackage{dcolumn}% Align table columns on decimal point
\usepackage{bm}% bold math
\usepackage{times}
\usepackage{multirow}
\usepackage{epstopdf}
\usepackage{color}
\usepackage[normalem]{ulem}
\usepackage{lineno}
\usepackage{xspace}

\definecolor{cmntblue}{rgb}{0.0, 0.58, 0.71}
\definecolor{cmntgreen}{rgb}{0.0, 0.42, 0.24}

\usepackage[draft, inline, nomargin]{fixme}
\fxsetup{theme=color, mode=multiuser}
\FXRegisterAuthor{bl}{1}{\color{blue}BL}  
\FXRegisterAuthor{pts}{anb}{\color{red}PTS}

\newcommand{\nuebar}{\ensuremath{\overline{\nu}_{e}}\xspace}
\newcommand{\uFive}{$^{235}$U\xspace}
\newcommand{\uEight}{$^{238}$U\xspace}
\newcommand{\pNine}{$^{239}$Pu\xspace}
\newcommand{\pOne}{$^{241}$Pu\xspace}
\newcommand{\ftimes}{$\times 10^{-43}$ cm$^2$/fission\xspace}

\begin{document}

\preprint{APS/123-QED}

\title{Diagnosing the Reactor Antineutrino Anomaly with Global Antineutrino Flux Data} 
\author{C. Giunti}%
\affiliation{INFN, Sezione di Torino, Via P. Giuria 1, I–10125 Torino, Italy}%

\author{Y. F. Li}%
\affiliation{Institute of High Energy Physics, Chinese Academy of Sciences, and School of Physical Sciences, University of Chinese Academy of Sciences, Beijing 100049, China}%

\author{B. R. Littlejohn}%
\affiliation{Physics Department, Illinois Institute of Technology, Chicago, IL 60616, USA}%

\author{P. T. Surukuchi}%
\affiliation{Physics Department, Illinois Institute of Technology, Chicago, IL 60616, USA}%

\date{\today}% It is always \today, today,

\begin{abstract}
We have examined the impact of new Daya Bay, Double Chooz, and RENO measurements on global fits of reactor antineutrino flux data to a variety of hypotheses regarding the origin of the reactor antineutrino anomaly.  
In comparing RENO and Daya Bay measurements of inverse beta decay (IBD) yield versus \pNine fission fraction, we find differing levels of precision in measurements of time-integrated yield and yield slope, but similar central values, leading to modestly enhanced isotopic IBD yield measurements in a joint fit of the two datasets.  
In the absence of sterile neutrino oscillations, global fits to all measurements now provide 3$\sigma$ preference for incorrect modelling of specific fission isotopes over common mis-modelling of all beta-converted isotopes.  
If sterile neutrino oscillations are considered, global IBD yield fits provide no substantial preference between oscillation-including and oscillation-excluding hypotheses: hybrid models containing both sterile neutrino oscillations and incorrect \uFive or \pNine flux predictions are favored at only 1-2$\sigma$ with respect to models where \uFive, \uEight, and \pNine are assumed to be incorrectly predicted.   

%Improved constraints on \uFive, \uEight, and \pNine IBD yields, as well as a 3$\sigma$ preference for a
%inverse beta-decay yields per fission provided by the inclusion of the new flux datasets are also reported.  
%using all available reactor flux data.  % and the spectral ratio measurements of DANSS, NEOS, and PROSPECT.
%The fits including both updated flux measurements and the spectral ratio data from DANSS, NEOS, and PROSPECT provide a preference for active-sterile neutrino oscillations similar to that previously reported using only DANSS and NEOS data.
%The combined fit of flux and spectral ratio data allows to obtain non-degenerate fits of both oscillation and flux parameters, and indicates a substantial preference in favor of hybrid models:
%in particular, a model containing sterile neutrinos and freely-fitted \uFive, \uEight, and \pNine~fluxes is favored at
%$\input{wrk/rea+evo+spe/mcm/235+238+239+OSC/tex/clv-235+238+239+OSC-OSC-sig.tex}\sigma$ over a model containing only sterile neutrino oscillations, and is favored at $\input{wrk/rea+evo+spe/mcm/235+238+239+OSC/tex/bef-clv-sig.tex}\sigma$ over a model containing only incorrect flux predictions. 
%Improved constraints on \uFive, \uEight, and \pNine

%\begin{description}
%\end{description}
\end{abstract}

\maketitle

%\tableofcontents

\section{Introduction}
\label{sec:introduction}

Within the particle physics community, there remains enduring interest in the observed deficit of detected reactor antineutrino (\nuebar) fluxes relative to the commonly-used conversion predictions~\cite{bib:huber,bib:mueller2011}.  
This deficit, called the reactor antineutrino anomaly~\cite{bib:mention2011,AnomalyWhite},
has been observed over a wide range of baselines and reactor fission fractions.

It has been hypothesized that the observed deficit could be the result of oscillation of reactor \nuebar into unobservable sterile types via one or more new mass-squared splittings of the order of 1~eV$^2$
(see the review in Ref.~\cite{Gariazzo:2015rra}).
Active-sterile oscillations should produce deficits in detected inverse beta decay (IBD) rates that are dependent on the baseline of the experiment and independent of the fission fractions in the observed reactors.  
An oscillation-based origin for the reactor antineutrino anomaly would have far-reaching experimental implications in neutrino physics, impacting the interpretation of prominent future long-baseline~\cite{deGouvea:2014aoa,Klop:2014ima,Kayser} and neutrinoless double beta decay experiments~\cite{Barry:2011wb,Li:2011ss,Girardi:2013zra,GiuntiDBD}.  %, as has been discussed in the literature~\cite{AnomalyWhite,Kayser,GiuntiDBD}.

The reactor antineutrino anomaly could also be caused by inaccuracies in the beta-converted \nuebar flux models of the fission isotopes \uFive, \pNine, and \pOne and the \textit{ab initio} model of \uEight~ 
\cite{Giunti:2016elf, hayes_first,hayes_shoulder,vogel_review,bib:dwyer, hayes_shape,surukuchi_flux,neutrons_shoulder,bib:HayesZ,bib:HayesMag}.  
%A number of papers have outlined possible inaccuracies in existing calculations~\cite{hayes_first,hayes_shoulder,vogel_review,bib:dwyer}; some of these have been further verified as capable of producing the observed anomalies~\cite{hayes_shape,surukuchi_flux}, while others have been ruled out as reasonable explanations~\cite{hayes_shape,neutrons_shoulder,bib:HayesZ,bib:HayesMag}.  
When converting measured fission beta spectra from the BILL spectrometer into attendant antineutrino spectra for \uFive, \pNine, and \pOne~\cite{bib:BILL,bib:ILL_1,bib:ILL_2,bib:ILL_3}, some inaccuracies could produce errors common to flux predictions of all isotopes: for example, non-consideration of important beta spectrum shape corrections~\cite{hayes_first}.   
On the other hand, other issues could produce errors specific to individual fission isotopes: for example, inconsistent calibration of neutron fluxes between different BILL beta spectrum measurements~\cite{bib:ILL_calib}.  %, or varying impacts of shape corrections in the conversion of different isotopes due to differing underlying fission yields.  
A model-based origin to the reactor anomaly should be reflected in a deficit in IBD detection rates that is not dependent on baseline and may or may not depend on the fission fractions of the experiment's reactor core.  

Hybrids of these two origins have also been highlighted in the literature~\cite{Giunti:2017yid,Gariazzo:2018mwd,schwetz_global}.
Such a scenario would produce dependencies of the measured IBD rate deficit on both fission fraction and baseline.  

Recent studies have analyzed the global IBD yield dataset to provide measurements of individual isotopic IBD yields and to assess the consistency of these datasets with hypotheses regarding the sources of the reactor flux anomaly.
Analyses including IBD yield measurements from highly \uFive-enriched (HEU) reactors provide indications that \uFive~predictions could be incorrect
\cite{Giunti:2016elf,Giunti:2017nww}, assuming the absence of active-sterile oscillations.
Yield measurements from periods of differing observed fission fractions from Daya Bay, termed its `flux evolution measurement', provide a distinct preference for incorrect \uFive~predictions over sterile neutrino oscillations as the sole cause of the anomaly~\cite{bib:prl_evol}.
Meanwhile, combined analyses of both Daya Bay evolution and global IBD yield measurements investigating a wider variety of hypotheses has shown that best fits to these data are produced by a hybrid of both incorrect flux predictions and active-sterile neutrino oscillations~\cite{Giunti:2017yid}.
Recent short-baseline measurements of the ratios of IBD energy spectra at different distances in the NEOS~\cite{bib:neos} and DANSS~\cite{danss_osc} experiments appear to also exhibit some preference for sterile neutrino oscillations~\cite{Gariazzo:2018mwd}, while other recent short-baseline measurements, PROSPECT~\cite{prospect_osc} and STEREO~\cite{stereo}, do not.  
%with a remarkable agreement on the values of the mixing parameters.
%The combination of these results with IBD yield results further strengthens indications in favor of the 'hybrid' hypothesis~\cite{Gariazzo:2018mwd}.

Recently, the community has seen the release of new results that are relevant to the investigation of these reactor anomaly hypotheses.  
In particular, the RENO collaboration has provided its first flux evolution measurement~\cite{RenoEvol}, and Daya Bay and Double Chooz have provided improved IBD yield measurements~\cite{bib:prd_flux,bib:dc_flux}.  
%and the PROSPECT~\cite{prospect_osc}
%and STEREO~\cite{Almazan:2018wln}
%collaborations have provided new multi-baseline relative spectrum distortion measurements at $<$~10~m reactor-detector distances.
The goal of this paper is to provide a comparison between Daya Bay and RENO flux evolution results, and to determine the impact of recent flux results on the global preference for the hybrid model of active-sterile oscillations and incorrect flux predictions.  
We find that RENO and Daya Bay results provide a generally consistent picture of reactor flux evolution, but differ in their precision and their ability to differentiate between sterile- and model-related deficit hypotheses.  
We also show that the addition of RENO and the new absolute flux results enables some improvement in the precision of isotopic IBD yield measurements.  
Finally, the global flux fits are found to exhibit only marginal preference for oscillation-including hypotheses over oscillation-excluding ones.  
%measurements have marginal present impact on the current global preference for non-zero sterile neutrino oscillations driven by the DANSS and NEOS short-baseline experiments.  

%\section{Updated IBD Yield Fits}
\section{Experimental Inputs}
\label{sec:Experimental}

Reactor antineutrino fluxes, sometimes reported experimentally as IBD yield, or the average flux times the IBD cross-section per fission, vary over time in a manner dependent on the fuel content of nearby reactor cores:
\begin{equation}\label{eq:Iso1}
\sigma_f(t) = \sum_i F_i(t) \sigma_i,
\end{equation}
where $\sigma_i$ is the IBD yield per fission for each parent fission isotope and $F_i(t)$ is the fission fraction of fission isotope $i$ in the measured reactor core
($i=5,8,9,1$ for \uFive, \uEight, \pNine, and \pOne, respectively).  
A number of experiments have provided single measurements of time-integral IBD yields; in this case the measured IBD yield $\bar{\sigma}_f$ is dictated by the average fission fractions of nearby reactor cores over the measurement time period.  
Some experiments have also provided multiple IBD yield measurements from different time periods of varying fission fraction; given the high degree of detector stability exhibited in these experiments, these measurements are highly systematically correlated.  
Using the measured $\sigma_f$ values and corresponding fission fractions, one can attempt to determine IBD yields for the individual fission isotopes, $\sigma_i$.

For the global IBD yield fits presented here, we use as input the existing body of measurements from Ref.~\cite{Gariazzo:2018mwd}, with a few exceptions.  
This includes time-integral measurements from ILL~\cite{bib:ILL_nu,ILL_nuFix}, Savannah River~\cite{bib:srp}, Krasnoyarsk~\cite{bib:Krasno1,bib:Krasno2,bib:Krasno4}, and Nucifer~\cite{bib:nucifer} at \uFive-burning HEU reactor cores, time-integral measurements from conventional low-enriched cores from Gosgen~\cite{bib:gosgen}, Rovno~\cite{bib:Rovno1,bib:Rovno2}, Bugey-3~\cite{bib:Bugey3_osc}, Bugey-4~\cite{bib:B4}, Palo Verde~\cite{bib:paloverde}, and Chooz~\cite{bib:chooz2}, and the flux evolution measurement of Daya Bay~\cite{bib:prl_evol}.  

In addition to these, we examine the inclusion of the new flux evolution measurement reported by the RENO collaboration~\cite{RenoEvol}, and the improved reactor flux measurements provided by Daya Bay~\cite{bib:prd_flux} and Double Chooz~\cite{bib:dc_flux}.  
RENO's new evolution result provides highly-correlated flux measurements at eight different  fission fraction values, while Daya Bay and Double Chooz flux measurements have been improved in precision to the 1.5\% and 1.0\% level.  
To account for a 0.3\% shift in the new Daya Bay time-integral flux with respect to the normalization of its older flux evolution result, all Daya Bay flux evolution data points are coherently shifted by this amount in our analysis.  
%The latter Double Chooz measurement represents the most precise reactor antineutrino flux measurement reported to date.  
%We note that the characteristics of flux evolution and normalization in this RENO reference have shifted somewhat over time; in the analysis we use the most recently-provided version of the RENO result.  

\begin{figure}[t]
\includegraphics[trim = 0.0cm 0.0cm 0.0cm 0.0cm, clip=true, width=0.49\textwidth]{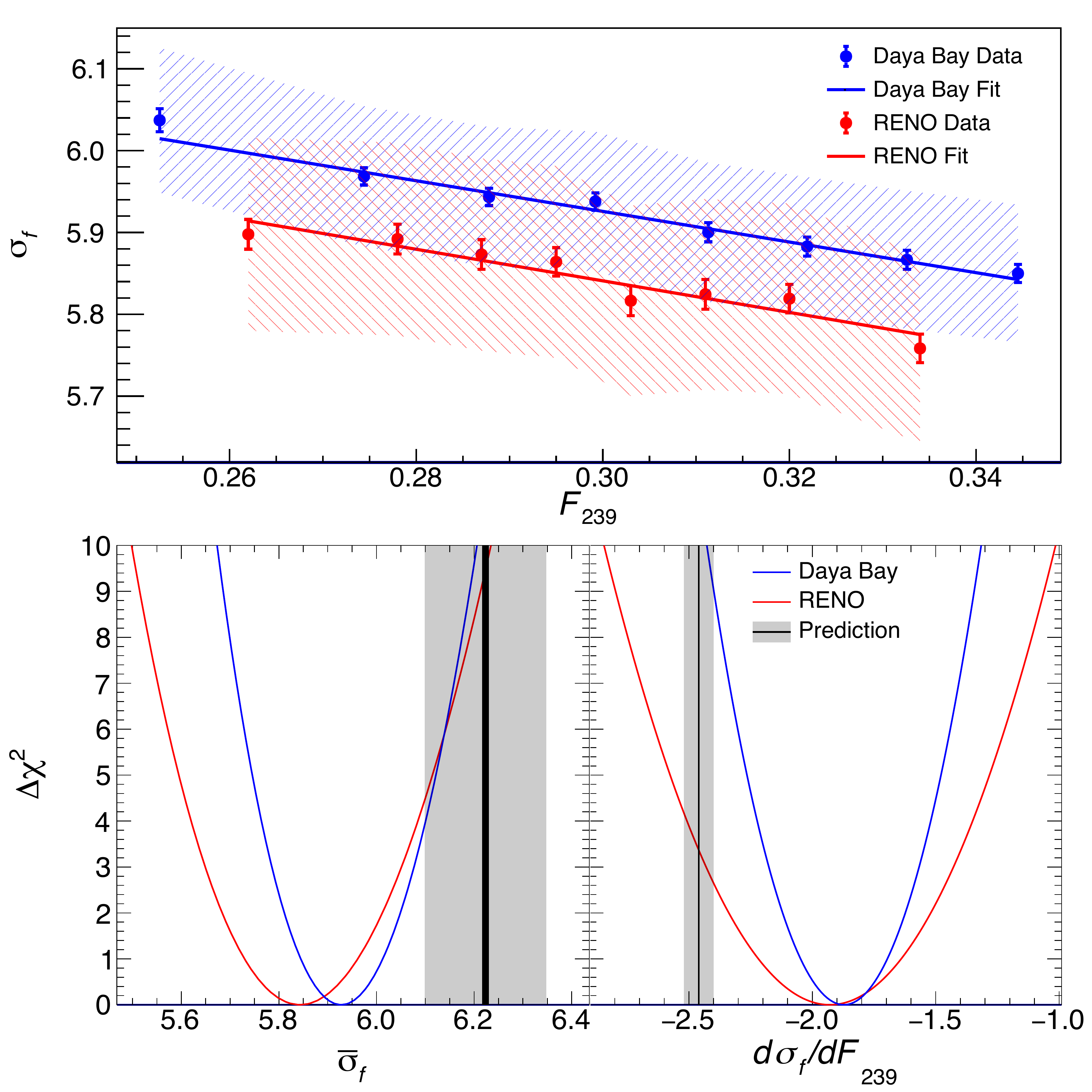}
\caption{Top: Daya Bay and RENO IBD yield measurements versus effective \pNine~fission fraction $F_{239}$. Error bars on each point represent statistical errors, while bands overlaying each dataset represent correlated uncertainties.  Bottom: Predicted and measured time-averaged IBD yields $\bar{\sigma}_{f}$ and yield slopes ($\frac{d\sigma_f}{dF_{239}}$).  A small predicted difference between Daya Bay and RENO $\bar{\sigma}_{f}$ due to differing average fission fractions is indicated by a thicker central value band. 
}
\label{fig:dr_data}
\end{figure}

To compare characteristics of the flux evolution data provided by Daya Bay and RENO, these results are overlaid in Fig.~\ref{fig:dr_data}.  
It can be seen that the two set of measurements span roughly equivalent fission fraction ranges and show similar-sized correlated uncertainty bands and uncorrelated statistical uncertainties.  
To further illustrate this comparison, we fit both experiments' data to linear functions as given in Ref.~\cite{bib:prl_evol}:
\begin{equation}\label{eq:Flux2}
\sigma_f(F_{239}) = \bar{\sigma}_f + \frac{d\sigma_f}{dF_{239}}
(F_{239}-\overline{F}_{239}),
%\overline{\sigma}_f = \sum_i \overline{F}_i \sigma_i,
\end{equation}
where $\bar{\sigma}_f$ is the time-integral IBD yield defined above, and $\frac{d\sigma_f}{dF_{239}}$~is the change in IBD yield per unit change in \pNine fission fraction $F_{239}$.  
Measured time integrated yields $\bar{\sigma}_f$ are (5.93$\pm$0.09) \ftimes and (5.84$\pm$0.12) \ftimes for Daya Bay and RENO, respectively, while measured slopes $\frac{d\sigma_f}{dF_{239}}$ are (-1.87$\pm$0.18) \ftimes and (-1.93$\pm$0.29) \ftimes.  
Yields and slopes are consistent within 1$\sigma$ between the two experiments.  
The modestly larger Daya Bay fission fraction range and smaller Daya Bay correlated (1.5\% versus 2.0\%) and uncorrelated ($\sim$0.1\% versus $\sim$0.2\% per data point) uncertainties produce smaller uncertainties in its measurement of the time-integral yield and the slope.  

Yield and slope values provided by the \uFive, \pNine, and \pOne~predictions of Ref.~\cite{bib:huber} and the \uEight~prediction of Ref~\cite{bib:mueller2011} are also pictured in Fig.~\ref{fig:dr_data}.  
The level of disagreement of Daya Bay and RENO with predicted time-integral yields, 2.0$\sigma$ and 2.2$\sigma$, respectively, are relatively similar.  
For relative slopes, $\frac{1}{\bar{\sigma}_f}\frac{d\sigma_f}{dF_{239}}$, Daya Bay and RENO show differing levels of consistency with predictions (3.1$\sigma$ versus 1.8$\sigma$, respectively), despite similar central values.  %; this is the product of the difference in precision in time-integral yield and slope measurements between experiments.  

\section{IBD Yield Fitting Procedure}
\label{sec:IBD}

To determine the best-fit isotopic IBD yields from the experimentally-provided IBD yields and fission fractions, we use the following $\chi^2$ definition: 
\begin{equation}
\begin{gathered}
\label{eq:Iso3}
\chi^2 = \sum_{a,b}\bigg(\sigma_{f,a} - P_{ee}^{a} \sum_i r_{i} F_{i,a} \sigma_i\bigg)
	\textrm{V}^{-1}_{ab} \\
	\times \bigg(\sigma_{f,b} - P_{ee}^{b} \sum_i r_{i} F_{i,b} \sigma_i\bigg) \\
     + \sum_{i,j}(\sigma^{th}_{i}-\sigma_{i}) \textrm{V}^{-1}_{\textrm{HM},ij}(\sigma^{th}_{j}-\sigma_{j}),
\end{gathered}
\end{equation}
where the experimental inputs $F_i$ and $\sigma_f$ are those described above,
with the indices $a$ and $b$ denoting the different experiments.
The covariance matrix $V_{ab}$ describing the uncertainties of the measured $\sigma_f$ values is based on the uncertainties provided in Refs.~\cite{Gariazzo:2017fdh,bib:prl_evol},
with alterations that take into account the new Daya Bay systematic uncertainty~\cite{bib:prd_flux}
and the uncertainties of the new RENO evolution data~\cite{RenoEvol}.  
Reduced fully-correlated systematic uncertainties in the new Daya Bay flux measurement are propagated to the flux evolution dataset via a proportional subtraction from all on- and off-diagonal elements of Daya Bay's uncertainty covariance matrix.  
The covariance matrix for the RENO evolution measurement is formed from the quoted statistics along the diagonal and a flat contribution to all elements from their quoted 2.1\% correlated systematic uncertainty.  
The isotopic IBD yields to be freely fitted are removed from the second expression, which constrains the isotopic IBD yields to the predicted yields $\sigma^{th}$ given in Refs.~\cite{bib:huber} for $\sigma^{th}_{5,9,1}$ and Ref.~\cite{bib:mueller2011} for $\sigma^{th}_{8}$,
with a theoretical uncertainty matrix $V_{\textrm{HM}}$ given in Table~3 of Ref.~\cite{Gariazzo:2017fdh}.
The primary fit parameters are the ratios $r_i$ between the best-fit and predicted yields,
and the neutrino mixing parameters
$\sin^22\vartheta_{ee} \equiv 4 |U_{e4}|^2 \left( 1 - |U_{e4}|^2 \right)$
and
$\Delta m^2_{41} \equiv m_{4}^2 - m_{1}^2$,
where $U$ is the neutrino mixing matrix and $m_{k}$ is the mass of the massive neutrino $\nu_{k}$,
which determine the averaged \nuebar oscillation survival probability
$P_{ee}^{a}$ for each experiment $a$ in the 3+1 neutrino mixing scheme.  
%(see the review in Ref.~\cite{Gariazzo:2015rra}).

In order to test a range of scenarios regarding the origin of the reactor flux anomaly, we apply a variety of constraints on the fit parameters $r_i$ and $P_{ee}$.  The first set of hypotheses assumes flux predictions are the sole origin of the flux anomaly; this is achieved by adding the constraint $P_{ee}$\,=\,1, as well as the following additional constraints on various isotopes' yields: 
\begin{itemize}
\item{ \textbf{235}: Constrain all $r_i$ except $r_5$.}
\item{ \textbf{239}: Constrain all $r_i$ except $r_9$.}
\item{ \textbf{235+239}: Constrain only $r_8$ and $r_1$.}
\item{ \textbf{235+238+239}: Constrain only $r_1$.}
\item{ \textbf{235=239=241+238}: Require common scaling of $r_{5,9,1}$; allow variations between  $r_{5,9,1}$ within the Huber uncorrelated uncertainties.}
\end{itemize}
Additional hypotheses including free fits of sterile neutrino oscillation parameters are also considered.  These scenarios correspond to the reactor flux anomaly being caused by sterile neutrino oscillations alone, or a hybrid combination of oscillations and incorrect flux modelling.
\begin{itemize}
\item{ \textbf{OSC}: Constrain all $r_i$ and freely fit $P_{ee}$.}
\item{ \textbf{239+OSC}: Same, but removing constraint on $r_{5}$.}
\item{ \textbf{235+OSC}: Same, but removing constraint on $r_{9}$.}
%\item{ \textbf{235+238+239+OSC}: Constrain only $r_1$.}
\end{itemize}

Rather than present best fits for all hypotheses and data combinations, we will highlight noteworthy results for each considered data combination.  

\section{Comparison of RENO and Daya Bay Results}
\label{sec:Comparison}

\begin{table*}[t]
\centering
\resizebox{\textwidth}{!}{
\begin{tabular}{c|ccc|ccc|ccc|}
&
\multicolumn{3}{c|}{\textbf{235}}
&
\multicolumn{3}{c|}{\textbf{239}}
&
\multicolumn{3}{c|}{\textbf{235+239}}
\\
&
Daya Bay
&
RENO
&
DB+RENO
&
Daya Bay
&
RENO
&
DB+RENO
&
Daya Bay
&
RENO
&
DB+RENO
\\
\hline
$\chi^{2}_{\text{min}}$ & $3.8
$                                                        & $5.2
$                                                        & $9.0
$                                                             & $10.2
$                                                        & $10.5
$                                                        & $16.1
$                                                             & $3.6
$                                                                    & $4.8
$                                                                    & $8.7
$
\\
$\text{NDF}$            & $7
$                                                        & $7
$                                                        & $15
$                                                             & $7
$                                                        & $7
$                                                        & $15
$                                                             & $6
$                                                                    & $6
$                                                                    & $14
$
\\
$\text{GoF}$            & $80
\%$                                                  & $64
\%$                                                  & $88
\%$                                                       & $18
\%$                                                  & $16
\%$                                                  & $38
\%$                                                       & $73
\%$                                                              & $58
\%$                                                              & $85
\%$
\\
$r_{5}$               & $0.927
\pm 0.016
$ & $0.917
\pm 0.026
$ & $0.925
\pm 0.015
$ & $(0.950
)$                                                  & $(0.961
)$                                                  & $(0.945
)$                                                       & $0.926
\pm 0.016$ & $0.913
\pm 0.027$ & $0.923
\pm 0.015$
\\
$r_{8}$               & $(0.988
)$                                                  & $(0.985
)$                                                  & $(0.985
)$                                                       & $(0.935
)$                                                  & $(0.949
)$                                                  & $(0.929
)$                                                       & $(1.001
)$                                                              & $(1.000
)$                                                              & $(1.003
)$
\\
$r_{9}$               & $(0.993
)$                                                  & $(0.989
)$                                                  & $(0.990
)$                                                       & $1.015
\pm 0.034
$ & $0.992
\pm 0.052
$ & $1.014
\pm 0.030
$ & $0.981
\pm 0.036$ & $0.957
\pm 0.054$ & $0.975
\pm 0.032$
\\
$r_{1}$               & $(0.994
)$                                                  & $(0.990
)$                                                  & $(0.991
)$                                                       & $(0.947
)$                                                  & $(0.959
)$                                                  & $(0.942
)$                                                       & $(1.001
)$                                                              & $(1.000
)$                                                              & $(1.000
)$
\end{tabular}
}
\caption{ Results of fitting Daya Bay and RENO flux evolution datasets with three  oscillation-excluding hypotheses in Section~\ref{sec:IBD} regarding the origin of the reactor anomaly.  Each hypothesis name denotes the unconstrained isotopic IBD yields in the fit.
The best-fit values and 1$\sigma$ ranges are given for unconstrained parameters, while the parenthetical values denote the best-fit values of the constrained fit parameters.  \label{tab1}
}
\end{table*}

\begin{table*}[t]
\centering
\resizebox{\textwidth}{!}{
\begin{tabular}{c|ccc|ccc|ccc|}
&
\multicolumn{3}{c|}{\textbf{OSC}}
&
\multicolumn{3}{c|}{\textbf{235+OSC}}
&
\multicolumn{3}{c|}{\textbf{239+OSC}}
\\
&
Daya Bay
&
RENO
&
DB+RENO
&
Daya Bay
&
RENO
&
DB+RENO
&
Daya Bay
&
RENO
&
DB+RENO
\\
\hline
$\chi^{2}_{\text{min}}$ & $ 9.5
$                                                & $6.5
$                                                & $16.3
$                                                     & $3.7
$                                                                    & $4.7
$                                                                     & $8.8
$                                                                         & $3.8
$                                                                    & $4.7
$                                                                    & $8.8
$
\\
$\text{NDF}$            & $7
$                                                & $7
$                                                & $15
$                                                     & $6
$                                                                    & $6
$                                                                     & $14
$                                                                         & $6
$                                                                    & $6
$                                                                    & $14
$
\\
$\text{GoF}$            & $22
\%$                                          & $48
\%$                                          & $37
\%$                                               & $72
\%$                                                              & $58
\%$                                                               & $85
\%$                                                                   & $70
\%$                                                              & $59
\%$                                                              & $84
\%$
\\
$P_{ee}$                & $0.944
\pm 0.025
$ & $0.935
\pm 0.030
$ & $0.939
\pm 0.024
$ & $0.988
\pm 0.021$ & $0.972
\pm 0.034$  & $0.986
\pm 0.022$ & $0.932
\pm 0.025$ & $0.921
\pm 0.031$ & $0.929
\pm 0.024$
\\
$r_{5}$               & $(1.003
)$                                          & $(1.001
)$                                          & $(1.004
)$                                               & $0.937
\pm 0.030$ & $0.943
\pm 0.047$  & $0.938
\pm 0.029$ & $(1.000
)$                                                              & $(1.000
)$                                                              & $(1.000
)$
\\
$r_{8}$               & $(1.037
)$                                          & $(1.009
)$                                          & $(1.045
)$                                               & $(1.004
)$                                                              & $(1.000
)$                                                               & $(1.001
)$                                                                   & $(1.003
)$                                                              & $(1.000
)$                                                              & $(1.002
)$
\\
$r_{9}$               & $(1.006
)$                                          & $(1.002
)$                                          & $(1.007
)$                                               & $(1.000
)$                                                              & $(1.000
)$                                                               & $(0.999
)$                                                                   & $1.094
\pm 0.048$ & $1.088
\pm 0.070$ & $1.094
\pm 0.043$
\\
$r_{1}$               & $(1.003
)$                                          & $(1.002
)$                                          & $(1.004
)$                                               & $(1.000
)$                                                              & $(1.000
)$                                                               & $(1.000
)$                                                                   & $(1.000
)$                                                              & $(1.000
)$                                                              & $(1.000
)$
\end{tabular}
}
\caption{ As Tab.~\ref{tab1}, for the three  oscillation-including hypotheses in Section~\ref{sec:IBD}.
%Results of fitting Daya Bay and RENO flux evolution datasets with three  oscillation-including hypotheses (Eq.~\ref{eq:Iso3}) regarding the origin of the reactor anomaly.  Each hypothesis name denotes the unconstrained parameters in the fit.
%The best-fit values and 1$\sigma$ ranges are given for unconstrained parameters, while the parenthetical values denote the best-fit values of the constrained fit parameters.
\label{tab2}
}
\end{table*}

\begin{figure}[t]
\includegraphics[clip=true, width=0.46\textwidth]{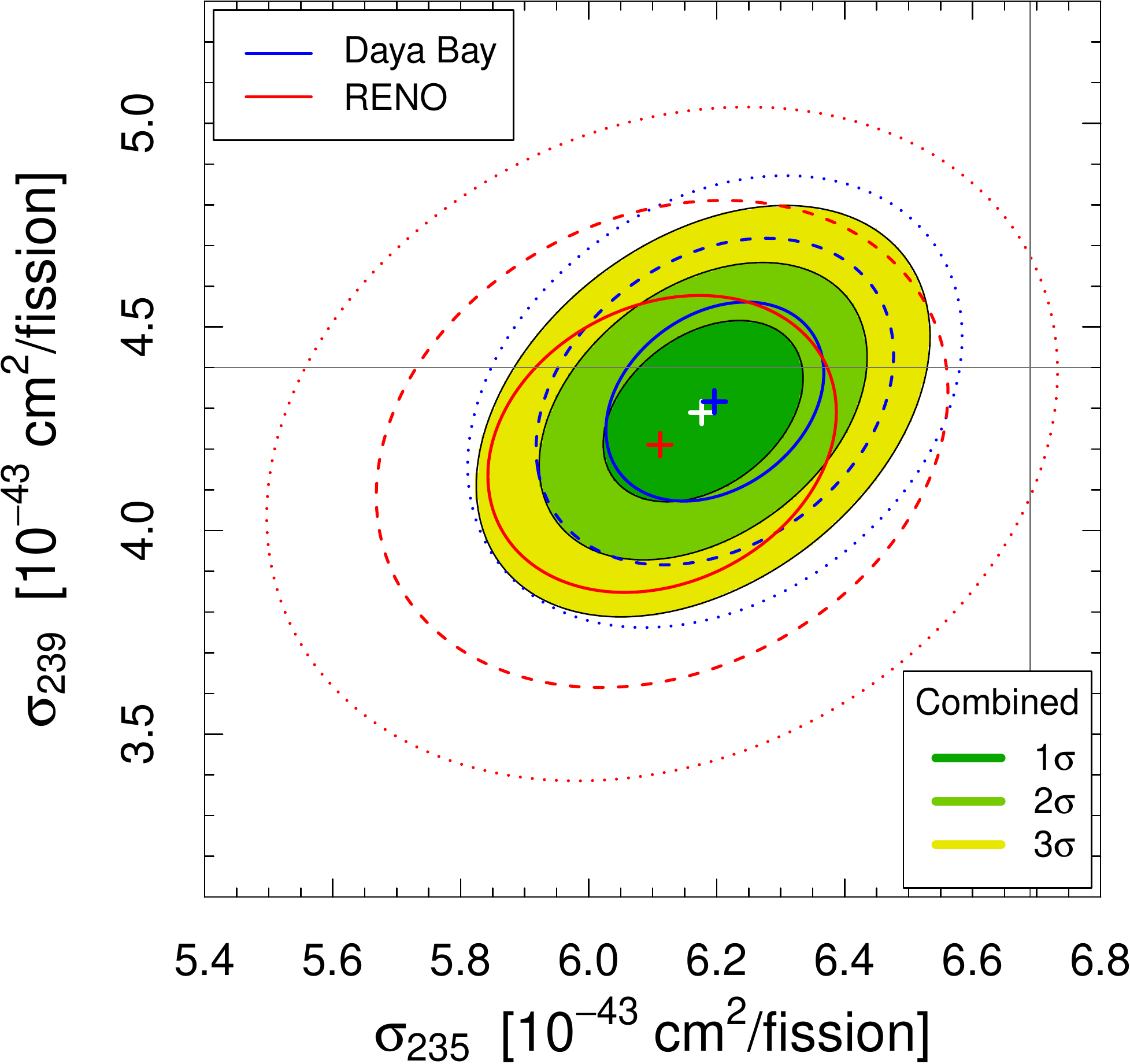}
\caption{Allowed regions for isotopic IBD yields of \uFive~and \pNine~provided by fits of updated Daya Bay and new RENO flux evolution results.}
\label{fig:dr_allowed}
\end{figure}

The allowed regions for the IBD yields of \uFive~and \pNine~in the absence of oscillations (\textbf{235+239} hypothesis) are shown in Fig.~\ref{fig:dr_allowed} for the new Daya Bay and RENO datasets, with best-fit values also overviewed in Tabs.~\ref{tab1} and~\ref{tab2}.  
Measured \uFive~IBD yields with respect to predictions are 0.926$\pm$0.016 and 0.913$\pm$0.027 for Daya Bay and RENO, respectively, while values for \pNine~are 0.981$\pm$0.036 and 0.957$\pm$0.054.  
The improvement in Daya Bay's detection efficiency has improved its IBD yield measurements~\cite{bib:prl_evol}: errors on \uFive~and \pNine~yields have been reduced by 0.9\% and 1.0\%, respectively, with respect to Ref.~\cite{Giunti:2017yid}, which uses identical $\chi^2$ definitions and theoretical IBD yield uncertainties.

RENO's best-fit \uFive~and \pNine IBD yields are quite similar to those obtained from the new Daya Bay dataset.  
Both experiments observe a substantial difference in \uFive~yield relative to Huber-Mueller, but not in \pNine~yield.  
The result of a combined fit of the Daya Bay and RENO datasets to the \textbf{235+239} hypothesis is also pictured in Fig.~\ref{fig:dr_allowed}.  
The combined fit produces minor improvements in precision for $r_5$ (from 1.6\% to 1.5\%) and $r_9$ (from 3.6\% to 3.2\%) over those obtained by Daya Bay alone.  
We note that we have not considered the \textbf{235+238+239} hypothesis here, as the combined RENO and Daya Bay data are not sufficient to constrain it.

\begin{table*}[t]
\centering
\begin{tabular}{c|cccccccc}
&
\textbf{235}
&
\textbf{239}
&
\textbf{235+239}
&
\textbf{235+238+239}
&
\textbf{235=239=241+238}
&
\textbf{OSC}
&
\textbf{235+OSC}
&
\textbf{239+OSC}
\\
\hline
$\chi^{2}_{\text{min}}$   & $34.6
$                                                    & $41.6
$                                                             & $34.1
$                                                                & $29.9
$                                                                            & $38.6
$                                                                                & $33.1$                                                     & $29.5$                                                                    & $26.9$                                                                   
\\
$\text{NDF}$              & $39
$                                                    & $39
$                                                             & $38
$                                                                & $37
$                                                                            & $38
$                                                                                & $38$                                                     & $37$                                                                    & $37$                                                                   
\\
$\text{GoF}$              & $67
\%$                                              & $36
\%$                                                       & $65
\%$                                                          & $79
\%$                                                                      & $44
\%$                                                                          & $69\%$                                               & $80\%$                                                              & $89\%$                                                             
\\
$r_{5}$                 & $0.933
\pm 0.010
$ & $(0.941
)$                                                       & $0.932
\pm 0.009$ & $0.952
\pm 0.014$ & $0.941
\pm 0.013$ & $(1.014)$                                                   & $0.984\pm 0.025$ & $(1.014)$                                                                 
\\
$r_{8}$                 & $(0.890
)$                                              & $(0.868
)$                                                       & $(0.914
)$                                                          & $0.672
\pm 0.135$ & $0.926
\pm 0.096$ & $(1.021)$                                                   & $(0.969)$                                                                  & $(0.956)$                                                                 
\\
$r_{9}$                 & $(0.987
)$                                              & $0.997
\pm 0.029
$          & $0.969
\pm 0.030$ & $1.042
\pm 0.046$ & $0.944
\pm 0.015$ & $(1.019)$                                                   & $(1.026)$                                                                  & $1.099\pm 0.040$
\\
$r_{1}$                 & $(0.989
)$                                              & $(0.938
)$                                                       & $(1.003
)$                                                          & $(1.001
)$                                                                      & $0.942
\pm 0.013$ & $(1.015)$                                                   & $(1.024)$                                                                  & $(1.015)$                                                                 
\\
$\Delta{m}^2_{41}$        &                                                                                              &                                                                                                       &                                                                                                              &                                                                                                                                      &                                                                                                                                              & $0.49^{+ 0.02 }_{- 0.03 }$ & $0.48^{+ 0.05 }_{- 0.03 }$            & $0.49\pm 0.02$           
\\
$\sin^2\!2\vartheta_{ee}$ &                                                                                              &                                                                                                       &                                                                                                              &                                                                                                                                      &                                                                                                                                              & $0.15\pm 0.04$ & $0.10^{+ 0.05 }_{- 0.04 }$            & $0.16\pm 0.04$           
\end{tabular}
\caption{ Results of fitting all time-integral and evolution flux measurements with hypotheses regarding the origin of the reactor anomaly.
Each hypothesis name denotes the unconstrained parameters in the fit.
The best-fit values and 1$\sigma$ ranges are given for unconstrained parameters, while the parenthetical values denote the best-fit values of the constrained fit parameters.
\label{tab:rea+evo}
}
\end{table*}

Results from fits of other hypotheses to the RENO and Daya Bay datasets are also overviewed in Tabs.~\ref{tab1} and \ref{tab2}\footnote{
In the analysis of Daya Bay and RENO evolution data alone we consider
the averaged survival probability
$ P_{ee} = 1 - \sin^22\vartheta_{ee} / 2 $,
because the source-detector distance is much larger than the oscillation length
for $ \Delta m^2_{41} \gtrsim 0.1 \, \text{eV}^2 $.
}.  
%Comparing $\chi^2_{\text{min}}$ between differing fit scenarios, we see that RENO has a more limited ability to differentiate between oscillation-based and model-based deficit origin hypotheses.  
While RENO data prefers hypotheses involving incorrect fluxes to the pure \textbf{OSC} hypothesis, all hypotheses in Tabs.~\ref{tab1} and \ref{tab2} save \textbf{239} exhibit a $\chi^2_{\text{min}}$ within 1.8 of the overall minimum.  
Meanwhile, Daya Bay data shows substantial preference for incorrect flux modelling: for example, a $\Delta \chi^2_{\text{min}}$ of
$5.7$
is seen between \textbf{OSC} and  \textbf{235} models.  
As discussed above, this difference in model discrimination power is due to the differences in experimental uncertainties between experiments, as opposed to substantial differences in the central values of best-fit parameters.  

Given the similarity of the best-fit parameters from the two datasets, a combined fit yields enhancements in the preferences against the \textbf{OSC} model.
The $\Delta \chi^2_{\text{min}}$ between \textbf{235} and \textbf{OSC} increased to from
$5.7$
for Daya Bay to
$7.3$
in the combined fit.
Using a frequentist Monte Carlo statistical analysis~\cite{Giunti:2017yid,surukuchi_flux}, this corresponds to a change in the preference of the \textbf{235} model against the \textbf{OSC} model from $2.6
\sigma$
for Daya Bay to
$2.9
\sigma$
in the combined fit of RENO and Daya Bay.

%Given the similarity of the best-fit parameters from the two datasets, a combined fit yields enhancements in the preferences against the \textbf{OSC} and \textbf{239} models.
%The $\Delta \chi^2_{\text{min}}$ between \textbf{235} and \textbf{OSC} (\textbf{239}) increased to from
%$\input{wrk/daya/uncstd/mcc-OSC-vs-235/delchi.dat}$
%($\input{wrk/daya/uncstd/mcc-239-vs-235/delchi.dat}$)
%for Daya Bay to
%$\input{wrk/daya+reno/uncstd/mcc-OSC-vs-235/delchi.dat}$
%($\input{wrk/daya+reno/uncstd/mcc-239-vs-235/delchi.dat}$)
%in the combined fit.
%Using a frequentist statistical analysis~\cite{Giunti:2017yid,surukuchi_flux}, this corresponds to a change in the preference of the \textbf{235} model against the \textbf{OSC} (\textbf{239}) model from %$\input{wrk/daya/uncstd/mcc-OSC-vs-235/dat/pvl-sig.dat}\sigma$
%($\input{wrk/daya/uncstd/mcc-239-vs-235/dat/pvl-sig.dat}\sigma$)
%for Daya Bay to
%$\input{wrk/daya+reno/uncstd/mcc-OSC-vs-235/dat/pvl-sig.dat}\sigma$
%($\input{wrk/daya+reno/uncstd/mcc-239-vs-235/dat/pvl-sig.dat}\sigma$)
%in the combined fit of RENO and Daya Bay.  

\section{Global Flux Fits}
\label{sec:Global}

We now turn to global fits of all time-integral and evolution IBD yield measurements.  
A comparison of global flux fit results for various oscillation-including or -excluding hypotheses introduced above are summarized in Tab.~\ref{tab:rea+evo}.

\begin{figure}[htb!]
\includegraphics[clip=true, width=0.49\textwidth]{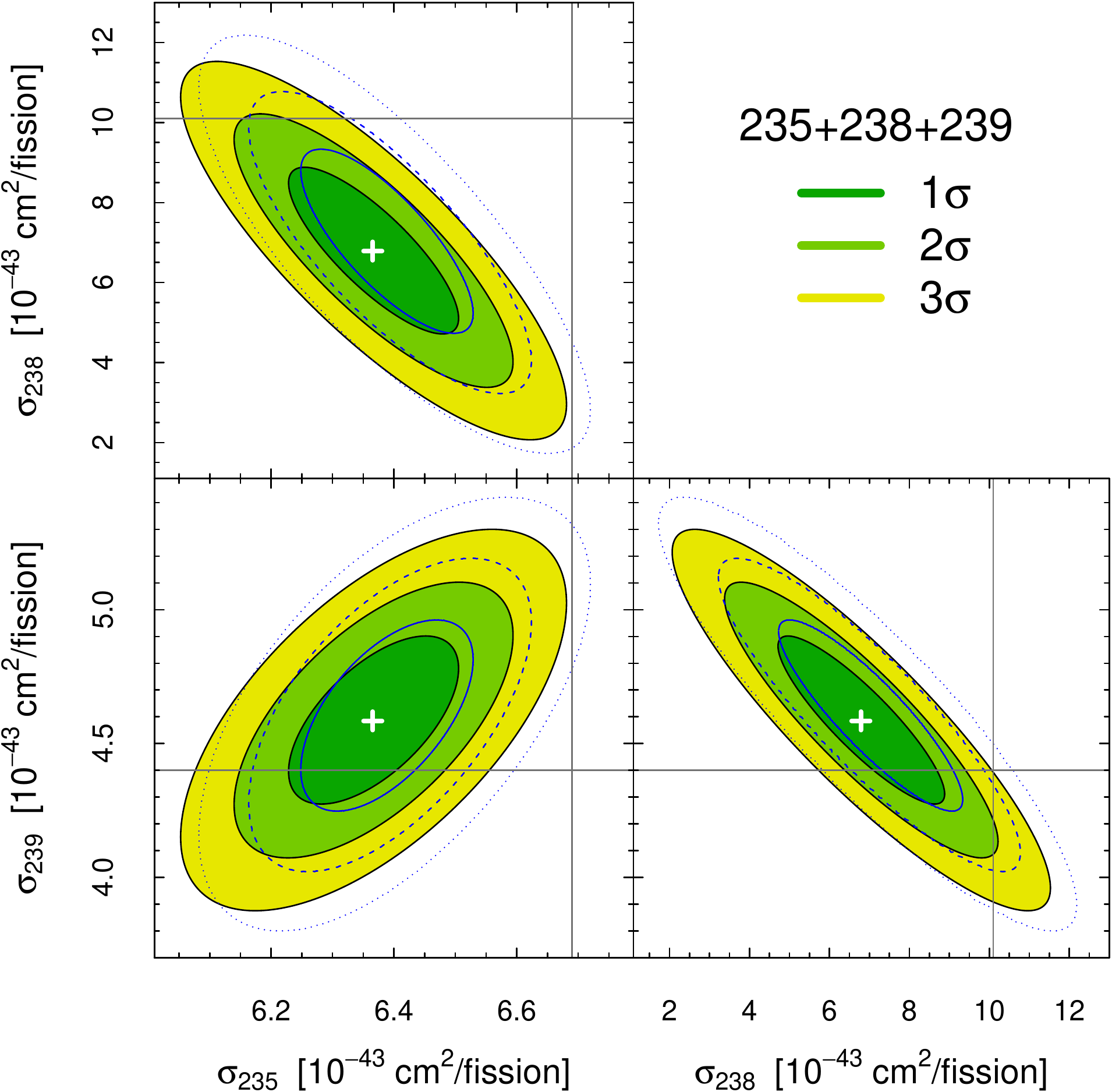}
\caption{Allowed regions for isotopic IBD yields of \uFive, \pNine, and \uEight provided by a fit of all IBD yield datasets to the \textbf{235+239+238} hypothesis described in the text.  The blue solid, dashed and dotted contours show the previous {235+239+238} regions obtained in Ref.~\cite{surukuchi_flux} at 1$\sigma$, 2$\sigma$, and 3$\sigma$, respectively. }
\label{fig:dr_threepanel}
\end{figure}

In the absence of oscillations, the allowed regions for the IBD yields of \uFive, \uEight, and \pNine (\textbf{235+239+238} hypothesis) are pictured in Fig,~\ref{fig:dr_threepanel},
along with the previously-determined best-fit values~\cite{surukuchi_flux}.
The best-fit $r$ values
$r_{5} = 0.952
\pm 0.014$,
$r_{8} = 0.672
\pm 0.135$, and
$r_{9} = 1.042
\pm 0.046$
are obtained for \uFive, \uEight, and \pNine, respectively.
The addition of improved Daya Bay, RENO and Double Chooz datasets has modestly improved the combined IBD yield constraints
for \uEight and \pNine,
that before were given by
$r_{8} = 0.695 \pm 0.163$
and
$r_{9} = 1.034 \pm 0.064$~\cite{surukuchi_flux},
whereas the uncertainty of the \uFive IBD yield is practically unchanged.
As in previous fits neglecting oscillations,
measured IBD yields for \uFive and \uEight disagree substantially with predicted central values,
now at the
$3.6\sigma$
and
$2.4\sigma$
level, while the yield for \pNine~remains consistent with its predicted value.

Comparing the different oscillation-excluding fits in Tab.~\ref{tab:rea+evo}, we find substantially higher $\chi^2$ values provided by the \textbf{235=239=241+238} and \textbf{239} hypotheses.  The former hypothesis corresponds to a common inaccuracy being present in all beta-conversion antineutrino flux predictions, while the latter corresponds to \pNine being the sole cause of the reactor flux anomaly.   
The overall worst fit is provided by the \textbf{239} hypothesis, which is in tension with existing flux constraints from \uFive-burning HEU reactors, as well as with the Daya Bay and RENO evolution datasets.  
If the 'common inaccuracy' \textbf{235=239=241+238} hypothesis is quantitatively compared with the hypothesis of uncorrelated inaccuracies between isotopes (\textbf{235+238+239} hypothesis) using a frequentist Monte Carlo statistical approach, the 8.7 $\Delta \chi^2$ between models corresponds to 3.0$\sigma$ preference for the latter hypothesis.  
Thus, if sterile neutrinos do not contribute to the reactor flux anomaly, the global IBD yield data favors model inaccuracies that are particular to specific fission isotopes.  
This conclusion is supported by recent work suggesting inconsistent calibration of neutron fluxes between fission beta spectrum measurements made by the BILL spectrometer~\cite{bib:ILL_calib}.  
In particular, the IBD yield data points to incorrect calibration of results for \uFive as well as incorrect \textit{ab initio} prediction of the \uEight flux.  

Considering now all possible oscillation-including or excluding hypotheses, we see that lowest $\chi^2_{\text{min}}$ values are delivered by the hybrid \textbf{235+OSC} and \textbf{239+OSC} models, as well by as the \textbf{235+238+239} model discussed above.  
Using the frequentist statistical approach, we find that the two hybrid models
\textbf{235+OSC} and \textbf{239+OSC}
are preferred at
$1.1
\sigma$
and
$1.8
\sigma$,
respectively,
to the most-preferred oscillation-excluding hypothesis
\textbf{235+238+239}.
%in spite of the small $\chi^2_{\text{min}}$ difference.  
The global flux fit thus does not provide definitive preference for or against the existence of sterile neutrino oscillations.
There is only a small improvement in comparing the two hybrid models
\textbf{235+OSC} and \textbf{239+OSC}
with the oscillation-excluding hypothesis \textbf{235+239},
which is disfavored at
$1.6
\sigma$
and
$2.3
\sigma$,
respectively,
or with the oscillation-excluding hypothesis \textbf{235},
which is disliked by the data at
$1.4
\sigma$
and
$2.0
\sigma$,
respectively.  
We also note that the \textbf{239+OSC} hypothesis is preferred to the \textbf{235+OSC} hypothesis by only
$1.5
\sigma$.  

\begin{figure*}[htb!]
\includegraphics[trim = 0.0cm 0.0cm 0.0cm 0.0cm, clip=true, width=0.9\textwidth]{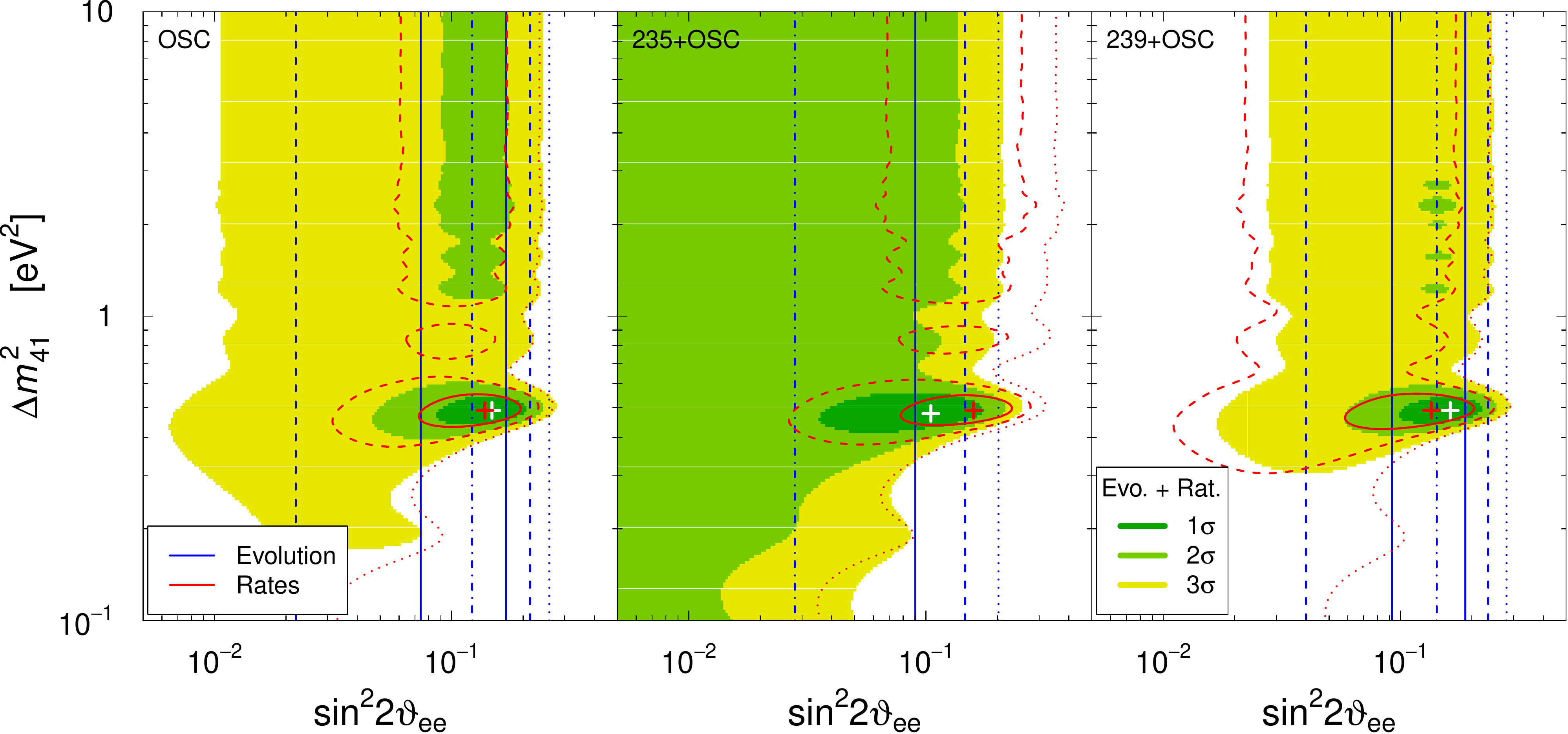}
\caption{1, 2, and 3$\sigma$ allowed oscillation parameter space regions for scenarios involving sterile neutrinos and incorrectly predicted fluxes from \uFive (center), \pNine (right), or neither isotope (left).
%$\chi^2_{\text{min}}$/NDF values are also cited for each scenario, along with p-values corresponding to preference for the best-fit of all flux models (\textbf{235+238+239+OSC}) over the pictured scenarios.
}
\label{fig:dr_oscflux}
\end{figure*}

To allow examination of oscillation-including hypotheses in more detail, the allowed regions in the plane of the oscillation parameters
$\Delta m^2_{41}$ and $\sin^22\vartheta_{ee}$
for the three oscillation-including hypotheses are pictured in Fig.~\ref{fig:dr_oscflux}.  
These hypotheses involving either zero or one incorrectly predicted fluxes provide similar best fit regions, except that these regions are shifted relatively between one another in $\sin^22\vartheta_{ee}$ space: \textbf{235+OSC} and \textbf{239+OSC} exhibit the lowest and highest best-fit $\sin^22\vartheta_{ee}$ values, respectively.  
This indicates that the null-oscillation IBD energy spectrum ratio results reported by DANSS, NEOS, PROSPECT, and STEREO are likely to have the most substantial impact on the oscillation parameter space suggested by the \textbf{239+OSC} hypothesis.  
%This distinction will become relevant when considering the combination of flux and spectral ratio results in the following section.  
We also note that although all three oscillation-including hypotheses fit the data well,
the hybrid \textbf{235+OSC} and \textbf{239+OSC} hypotheses are preferred to the pure oscillation hypothesis (\textbf{OSC}) by
$1.9\sigma$ and $2.5\sigma$, respectively.

\section{Summary}
\label{sec:Summary}

We have performed global fits of the complete set of reactor \nuebar flux data, including new flux evolution data from RENO and new time-integrated flux measurements from Daya Bay and Double Chooz.  
We find that the RENO and Daya Bay flux evolution datasets are similar in their preferred central values of absolute IBD yield and yield slopes,  but differ in their level of precision, due to lower statistical and systematic uncertainties and wider fission fraction ranges provided by Daya Bay.  
A joint fit of the two datasets leads to an increased 2.9$\sigma$ preference for incorrect \uFive predictions over sterile neutrino oscillations as the sole source of the reactor antineutrino flux anomaly.  
%, this precision difference leads Daya Bay data to be dominant in the determination of IBD yields, and in the generation of preference for specific hypotheses regarding the origin of the reactor flux anomaly.  
A global fit of these two evolution results and all time-integrated \nuebar flux measurements produces improved IBD yield constraints over those reported in previous publications.  
We find that all \nuebar flux data, in the absence of oscillations, now disfavor a common inaccuracy among all beta conversion predictions at 3.0$\sigma$ confidence level.   
We also find that flux data, alone, currently does not provide a sizable preference for  oscillation-including or oscillation-excluding hypotheses.

\begin{acknowledgments}
We would like to thank Yonas Gebre, Soo-Bong Kim, Marco Laveder, and Xianyi Zhang for useful discussions.
The work of Y.F. Li was supported in part by the National Natural Science Foundation of China under Grant No. 11835013, by the Strategic Priority Research Program of the Chinese Academy of Sciences under Grant No.~XDA10010100, and by the CAS Center for Excellence in Particle Physics (CCEPP).
The work of B.R. Littlejohn and P. T. Surukuchi was supported by the DOE Office of Science, under award No. DE-SC0008347.
\end{acknowledgments}

\bibliographystyle{apsrev4-1}
\bibliography{refs}{}

\begin{thebibliography}{54}%
\makeatletter
\providecommand \@ifxundefined [1]{%
 \@ifx{#1\undefined}
}%
\providecommand \@ifnum [1]{%
 \ifnum #1\expandafter \@firstoftwo
 \else \expandafter \@secondoftwo
 \fi
}%
\providecommand \@ifx [1]{%
 \ifx #1\expandafter \@firstoftwo
 \else \expandafter \@secondoftwo
 \fi
}%
\providecommand \natexlab [1]{#1}%
\providecommand \enquote  [1]{``#1''}%
\providecommand \bibnamefont  [1]{#1}%
\providecommand \bibfnamefont [1]{#1}%
\providecommand \citenamefont [1]{#1}%
\providecommand \href@noop [0]{\@secondoftwo}%
\providecommand \href [0]{\begingroup \@sanitize@url \@href}%
\providecommand \@href[1]{\@@startlink{#1}\@@href}%
\providecommand \@@href[1]{\endgroup#1\@@endlink}%
\providecommand \@sanitize@url [0]{\catcode `\\12\catcode `\$12\catcode
  `\&12\catcode `\#12\catcode `\^12\catcode `\_12\catcode `\%12\relax}%
\providecommand \@@startlink[1]{}%
\providecommand \@@endlink[0]{}%
\providecommand \url  [0]{\begingroup\@sanitize@url \@url }%
\providecommand \@url [1]{\endgroup\@href {#1}{\urlprefix }}%
\providecommand \urlprefix  [0]{URL }%
\providecommand \Eprint [0]{\href }%
\providecommand \doibase [0]{http://dx.doi.org/}%
\providecommand \selectlanguage [0]{\@gobble}%
\providecommand \bibinfo  [0]{\@secondoftwo}%
\providecommand \bibfield  [0]{\@secondoftwo}%
\providecommand \translation [1]{[#1]}%
\providecommand \BibitemOpen [0]{}%
\providecommand \bibitemStop [0]{}%
\providecommand \bibitemNoStop [0]{.\EOS\space}%
\providecommand \EOS [0]{\spacefactor3000\relax}%
\providecommand \BibitemShut  [1]{\csname bibitem#1\endcsname}%
\let\auto@bib@innerbib\@empty
%</preamble>
\bibitem [{\citenamefont {Huber}(2011)}]{bib:huber}%
  \BibitemOpen
  \bibfield  {author} {\bibinfo {author} {\bibfnamefont {P.}~\bibnamefont
  {Huber}},\ }\href {\doibase 10.1103/PhysRevC.85.029901,
  10.1103/PhysRevC.84.024617} {\bibfield  {journal} {\bibinfo  {journal} {Phys.
  Rev. C}\ }\textbf {\bibinfo {volume} {84}},\ \bibinfo {pages} {024617}
  (\bibinfo {year} {2011})}\BibitemShut {NoStop}%
\bibitem [{\citenamefont {Mueller}\ \emph {et~al.}(2011)\citenamefont {Mueller}
  \emph {et~al.}}]{bib:mueller2011}%
  \BibitemOpen
  \bibfield  {author} {\bibinfo {author} {\bibfnamefont {T.~A.}\ \bibnamefont
  {Mueller}} \emph {et~al.},\ }\href {\doibase 10.1103/PhysRevC.83.054615}
  {\bibfield  {journal} {\bibinfo  {journal} {Phys. Rev. C}\ }\textbf {\bibinfo
  {volume} {83}},\ \bibinfo {pages} {054615} (\bibinfo {year}
  {2011})}\BibitemShut {NoStop}%
\bibitem [{\citenamefont {Mention}\ \emph {et~al.}(2011)\citenamefont {Mention}
  \emph {et~al.}}]{bib:mention2011}%
  \BibitemOpen
  \bibfield  {author} {\bibinfo {author} {\bibfnamefont {G.}~\bibnamefont
  {Mention}} \emph {et~al.},\ }\href {\doibase 10.1103/PhysRevD.83.073006}
  {\bibfield  {journal} {\bibinfo  {journal} {Phys. Rev. D}\ }\textbf {\bibinfo
  {volume} {83}},\ \bibinfo {pages} {073006} (\bibinfo {year}
  {2011})}\BibitemShut {NoStop}%
\bibitem [{\citenamefont {Abazajian}\ \emph {et~al.}(2012)\citenamefont
  {Abazajian} \emph {et~al.}}]{AnomalyWhite}%
  \BibitemOpen
  \bibfield  {author} {\bibinfo {author} {\bibfnamefont {K.}~\bibnamefont
  {Abazajian}} \emph {et~al.},\ }\href@noop {} {\  (\bibinfo {year} {2012})},\
  \Eprint {http://arxiv.org/abs/1204.5379} {arXiv:1204.5379 [hep-ph]}
  \BibitemShut {NoStop}%
\bibitem [{\citenamefont {Gariazzo}\ \emph {et~al.}(2016)\citenamefont
  {Gariazzo}, \citenamefont {Giunti}, \citenamefont {Laveder}, \citenamefont
  {Li},\ and\ \citenamefont {Zavanin}}]{Gariazzo:2015rra}%
  \BibitemOpen
  \bibfield  {author} {\bibinfo {author} {\bibfnamefont {S.}~\bibnamefont
  {Gariazzo}}, \bibinfo {author} {\bibfnamefont {C.}~\bibnamefont {Giunti}},
  \bibinfo {author} {\bibfnamefont {M.}~\bibnamefont {Laveder}}, \bibinfo
  {author} {\bibfnamefont {Y.~F.}\ \bibnamefont {Li}}, \ and\ \bibinfo {author}
  {\bibfnamefont {E.~M.}\ \bibnamefont {Zavanin}},\ }\href {\doibase
  10.1088/0954-3899/43/3/033001} {\bibfield  {journal} {\bibinfo  {journal} {J.
  Phys.}\ }\textbf {\bibinfo {volume} {G43}},\ \bibinfo {pages} {033001}
  (\bibinfo {year} {2016})},\ \Eprint {http://arxiv.org/abs/1507.08204}
  {arXiv:1507.08204 [hep-ph]} \BibitemShut {NoStop}%
%%CITATION = ARXIV:1507.08204;%%
\bibitem [{\citenamefont {de~GouvÃªa}\ \emph {et~al.}(2015)\citenamefont
  {de~GouvÃªa}, \citenamefont {Kelly},\ and\ \citenamefont
  {Kobach}}]{deGouvea:2014aoa}%
  \BibitemOpen
  \bibfield  {author} {\bibinfo {author} {\bibfnamefont {A.}~\bibnamefont
  {de~GouvÃªa}}, \bibinfo {author} {\bibfnamefont {K.~J.}\ \bibnamefont
  {Kelly}}, \ and\ \bibinfo {author} {\bibfnamefont {A.}~\bibnamefont
  {Kobach}},\ }\href {\doibase 10.1103/PhysRevD.91.053005} {\bibfield
  {journal} {\bibinfo  {journal} {Phys. Rev.}\ }\textbf {\bibinfo {volume}
  {D91}},\ \bibinfo {pages} {053005} (\bibinfo {year} {2015})},\ \Eprint
  {http://arxiv.org/abs/1412.1479} {arXiv:1412.1479 [hep-ph]} \BibitemShut
  {NoStop}%
%%CITATION = ARXIV:1412.1479;%%
\bibitem [{\citenamefont {Klop}\ and\ \citenamefont
  {Palazzo}(2015)}]{Klop:2014ima}%
  \BibitemOpen
  \bibfield  {author} {\bibinfo {author} {\bibfnamefont {N.}~\bibnamefont
  {Klop}}\ and\ \bibinfo {author} {\bibfnamefont {A.}~\bibnamefont {Palazzo}},\
  }\href {\doibase 10.1103/PhysRevD.91.073017} {\bibfield  {journal} {\bibinfo
  {journal} {Phys. Rev.}\ }\textbf {\bibinfo {volume} {D91}},\ \bibinfo {pages}
  {073017} (\bibinfo {year} {2015})},\ \Eprint {http://arxiv.org/abs/1412.7524}
  {arXiv:1412.7524 [hep-ph]} \BibitemShut {NoStop}%
%%CITATION = ARXIV:1412.7524;%%
\bibitem [{\citenamefont {Gandhi}\ \emph {et~al.}(2015)\citenamefont {Gandhi},
  \citenamefont {Kayser}, \citenamefont {Masud},\ and\ \citenamefont
  {Prakash}}]{Kayser}%
  \BibitemOpen
  \bibfield  {author} {\bibinfo {author} {\bibfnamefont {R.}~\bibnamefont
  {Gandhi}}, \bibinfo {author} {\bibfnamefont {B.}~\bibnamefont {Kayser}},
  \bibinfo {author} {\bibfnamefont {M.}~\bibnamefont {Masud}}, \ and\ \bibinfo
  {author} {\bibfnamefont {S.}~\bibnamefont {Prakash}},\ }\href@noop {}
  {\bibfield  {journal} {\bibinfo  {journal} {JHEP}\ }\textbf {\bibinfo
  {volume} {1511:039}} (\bibinfo {year} {2015})}\BibitemShut {NoStop}%
\bibitem [{\citenamefont {Barry}\ \emph {et~al.}(2011)\citenamefont {Barry},
  \citenamefont {Rodejohann},\ and\ \citenamefont {Zhang}}]{Barry:2011wb}%
  \BibitemOpen
  \bibfield  {author} {\bibinfo {author} {\bibfnamefont {J.}~\bibnamefont
  {Barry}}, \bibinfo {author} {\bibfnamefont {W.}~\bibnamefont {Rodejohann}}, \
  and\ \bibinfo {author} {\bibfnamefont {H.}~\bibnamefont {Zhang}},\ }\href
  {\doibase 10.1007/JHEP07(2011)091} {\bibfield  {journal} {\bibinfo  {journal}
  {JHEP}\ }\textbf {\bibinfo {volume} {07}},\ \bibinfo {pages} {091} (\bibinfo
  {year} {2011})},\ \Eprint {http://arxiv.org/abs/1105.3911} {arXiv:1105.3911
  [hep-ph]} \BibitemShut {NoStop}%
%%CITATION = ARXIV:1105.3911;%%
\bibitem [{\citenamefont {Li}\ and\ \citenamefont {Liu}(2012)}]{Li:2011ss}%
  \BibitemOpen
  \bibfield  {author} {\bibinfo {author} {\bibfnamefont {Y.~F.}\ \bibnamefont
  {Li}}\ and\ \bibinfo {author} {\bibfnamefont {S.-s.}\ \bibnamefont {Liu}},\
  }\href {\doibase 10.1016/j.physletb.2011.11.054} {\bibfield  {journal}
  {\bibinfo  {journal} {Phys. Lett. B}\ }\textbf {\bibinfo {volume} {706}},\
  \bibinfo {pages} {406} (\bibinfo {year} {2012})},\ \Eprint
  {http://arxiv.org/abs/1110.5795} {arXiv:1110.5795 [hep-ph]} \BibitemShut
  {NoStop}%
%%CITATION = ARXIV:1110.5795;%%
\bibitem [{\citenamefont {Girardi}\ \emph {et~al.}(2013)\citenamefont
  {Girardi}, \citenamefont {Meroni},\ and\ \citenamefont
  {Petcov}}]{Girardi:2013zra}%
  \BibitemOpen
  \bibfield  {author} {\bibinfo {author} {\bibfnamefont {I.}~\bibnamefont
  {Girardi}}, \bibinfo {author} {\bibfnamefont {A.}~\bibnamefont {Meroni}}, \
  and\ \bibinfo {author} {\bibfnamefont {S.~T.}\ \bibnamefont {Petcov}},\
  }\href {\doibase 10.1007/JHEP11(2013)146} {\bibfield  {journal} {\bibinfo
  {journal} {JHEP}\ }\textbf {\bibinfo {volume} {11}},\ \bibinfo {pages} {146}
  (\bibinfo {year} {2013})},\ \Eprint {http://arxiv.org/abs/1308.5802}
  {arXiv:1308.5802 [hep-ph]} \BibitemShut {NoStop}%
%%CITATION = ARXIV:1308.5802;%%
\bibitem [{\citenamefont {Giunti}\ and\ \citenamefont
  {Zavanin}(2015)}]{GiuntiDBD}%
  \BibitemOpen
  \bibfield  {author} {\bibinfo {author} {\bibfnamefont {C.}~\bibnamefont
  {Giunti}}\ and\ \bibinfo {author} {\bibfnamefont {E.~M.}\ \bibnamefont
  {Zavanin}},\ }\href {\doibase 10.1007/JHEP07(2015)171} {\bibfield  {journal}
  {\bibinfo  {journal} {JHEP}\ }\textbf {\bibinfo {volume} {07}},\ \bibinfo
  {pages} {171} (\bibinfo {year} {2015})},\ \Eprint
  {http://arxiv.org/abs/1505.00978} {arXiv:1505.00978 [hep-ph]} \BibitemShut
  {NoStop}%
%%CITATION = ARXIV:1505.00978;%%
\bibitem [{\citenamefont {Giunti}(2017{\natexlab{a}})}]{Giunti:2016elf}%
  \BibitemOpen
  \bibfield  {author} {\bibinfo {author} {\bibfnamefont {C.}~\bibnamefont
  {Giunti}},\ }\href {\doibase 10.1016/j.physletb.2016.11.028} {\bibfield
  {journal} {\bibinfo  {journal} {Phys. Lett. B}\ }\textbf {\bibinfo {volume}
  {764}},\ \bibinfo {pages} {145} (\bibinfo {year} {2017}{\natexlab{a}})},\
  \Eprint {http://arxiv.org/abs/1608.04096} {arXiv:1608.04096 [hep-ph]}
  \BibitemShut {NoStop}%
%%CITATION = ARXIV:1608.04096;%%
\bibitem [{\citenamefont {{A.~Hayes, J. Friar, G. Garvey, G. Jungman, and G.
  Jonkmans}}(2014)}]{hayes_first}%
  \BibitemOpen
  \bibfield  {author} {\bibinfo {author} {\bibnamefont {{A.~Hayes, J. Friar, G.
  Garvey, G. Jungman, and G. Jonkmans}}},\ }\href {\doibase
  10.1103/PhysRevLett.112.202501} {\bibfield  {journal} {\bibinfo  {journal}
  {Phys. Rev. Lett.}\ }\textbf {\bibinfo {volume} {112}},\ \bibinfo {pages}
  {202501} (\bibinfo {year} {2014})}\BibitemShut {NoStop}%
\bibitem [{\citenamefont {Hayes}\ \emph {et~al.}(2015)\citenamefont {Hayes},
  \citenamefont {Friar}, \citenamefont {Garvey}, \citenamefont {Ibeling},
  \citenamefont {Jungman}, \citenamefont {Kawano},\ and\ \citenamefont
  {Mills}}]{hayes_shoulder}%
  \BibitemOpen
  \bibfield  {author} {\bibinfo {author} {\bibfnamefont {A.}~\bibnamefont
  {Hayes}}, \bibinfo {author} {\bibfnamefont {J.}~\bibnamefont {Friar}},
  \bibinfo {author} {\bibfnamefont {G.}~\bibnamefont {Garvey}}, \bibinfo
  {author} {\bibfnamefont {D.}~\bibnamefont {Ibeling}}, \bibinfo {author}
  {\bibfnamefont {G.}~\bibnamefont {Jungman}}, \bibinfo {author} {\bibfnamefont
  {T.}~\bibnamefont {Kawano}}, \ and\ \bibinfo {author} {\bibfnamefont
  {R.}~\bibnamefont {Mills}},\ }\href {\doibase 10.1103/PhysRevD.92.033015}
  {\bibfield  {journal} {\bibinfo  {journal} {Phys. Rev. D}\ }\textbf {\bibinfo
  {volume} {92}},\ \bibinfo {pages} {033015} (\bibinfo {year}
  {2015})}\BibitemShut {NoStop}%
\bibitem [{\citenamefont {Hayes}\ and\ \citenamefont
  {Vogel}(2016)}]{vogel_review}%
  \BibitemOpen
  \bibfield  {author} {\bibinfo {author} {\bibfnamefont {A.}~\bibnamefont
  {Hayes}}\ and\ \bibinfo {author} {\bibfnamefont {P.}~\bibnamefont {Vogel}},\
  }\href@noop {} {\bibfield  {journal} {\bibinfo  {journal} {Ann. Rev. Nucl.
  Part. Sci.}\ }\textbf {\bibinfo {volume} {66}},\ \bibinfo {pages} {219}
  (\bibinfo {year} {2016})}\BibitemShut {NoStop}%
\bibitem [{\citenamefont {Dwyer}\ and\ \citenamefont
  {Langford}(2015)}]{bib:dwyer}%
  \BibitemOpen
  \bibfield  {author} {\bibinfo {author} {\bibfnamefont {D.~A.}\ \bibnamefont
  {Dwyer}}\ and\ \bibinfo {author} {\bibfnamefont {T.~J.}\ \bibnamefont
  {Langford}},\ }\href {\doibase 10.1103/PhysRevLett.114.012502} {\bibfield
  {journal} {\bibinfo  {journal} {Phys. Rev. Lett.}\ }\textbf {\bibinfo
  {volume} {114}},\ \bibinfo {pages} {012502} (\bibinfo {year}
  {2015})}\BibitemShut {NoStop}%
\bibitem [{\citenamefont {Hayes}\ \emph {et~al.}(2017)\citenamefont {Hayes},
  \citenamefont {McCutchan},\ and\ \citenamefont {Sonzogni}}]{hayes_shape}%
  \BibitemOpen
  \bibfield  {author} {\bibinfo {author} {\bibfnamefont {A.~C.}\ \bibnamefont
  {Hayes}}, \bibinfo {author} {\bibfnamefont {E.~A.}\ \bibnamefont
  {McCutchan}}, \ and\ \bibinfo {author} {\bibfnamefont {A.~A.}\ \bibnamefont
  {Sonzogni}},\ }\href@noop {} {\bibfield  {journal} {\bibinfo  {journal}
  {Phys. Rev. Lett.}\ }\textbf {\bibinfo {volume} {119}},\ \bibinfo {pages}
  {112501} (\bibinfo {year} {2017})}\BibitemShut {NoStop}%
\bibitem [{\citenamefont {Gebre}\ \emph {et~al.}(2018)\citenamefont {Gebre},
  \citenamefont {Littlejohn},\ and\ \citenamefont
  {Surukuchi}}]{surukuchi_flux}%
  \BibitemOpen
  \bibfield  {author} {\bibinfo {author} {\bibfnamefont {Y.}~\bibnamefont
  {Gebre}}, \bibinfo {author} {\bibfnamefont {B.}~\bibnamefont {Littlejohn}}, \
  and\ \bibinfo {author} {\bibfnamefont {P.~T.}\ \bibnamefont {Surukuchi}},\
  }\href@noop {} {\bibfield  {journal} {\bibinfo  {journal} {Phys. Rev. D}\
  }\textbf {\bibinfo {volume} {97}},\ \bibinfo {pages} {013003} (\bibinfo
  {year} {2018})}\BibitemShut {NoStop}%
\bibitem [{\citenamefont {Littlejohn}\ \emph {et~al.}(2018)\citenamefont
  {Littlejohn}, \citenamefont {Conant}, \citenamefont {Dwyer}, \citenamefont
  {Erickson}, \citenamefont {Gustafson},\ and\ \citenamefont
  {Hermanek}}]{neutrons_shoulder}%
  \BibitemOpen
  \bibfield  {author} {\bibinfo {author} {\bibfnamefont {B.~R.}\ \bibnamefont
  {Littlejohn}}, \bibinfo {author} {\bibfnamefont {A.}~\bibnamefont {Conant}},
  \bibinfo {author} {\bibfnamefont {D.~A.}\ \bibnamefont {Dwyer}}, \bibinfo
  {author} {\bibfnamefont {A.}~\bibnamefont {Erickson}}, \bibinfo {author}
  {\bibfnamefont {I.}~\bibnamefont {Gustafson}}, \ and\ \bibinfo {author}
  {\bibfnamefont {K.}~\bibnamefont {Hermanek}},\ }\href@noop {} {\bibfield
  {journal} {\bibinfo  {journal} {Phys. Rev. D}\ }\textbf {\bibinfo {volume}
  {97}},\ \bibinfo {pages} {073007} (\bibinfo {year} {2018})}\BibitemShut
  {NoStop}%
\bibitem [{\citenamefont {Wang}\ and\ \citenamefont
  {Hayes}(2017)}]{bib:HayesZ}%
  \BibitemOpen
  \bibfield  {author} {\bibinfo {author} {\bibfnamefont {X.}~\bibnamefont
  {Wang}}\ and\ \bibinfo {author} {\bibfnamefont {A.}~\bibnamefont {Hayes}},\
  }\href@noop {} {\bibfield  {journal} {\bibinfo  {journal} {Phys. Rev. C}\
  }\textbf {\bibinfo {volume} {95}},\ \bibinfo {pages} {064313} (\bibinfo
  {year} {2017})}\BibitemShut {NoStop}%
\bibitem [{\citenamefont {Wang}\ \emph {et~al.}(2016)\citenamefont {Wang},
  \citenamefont {Friar},\ and\ \citenamefont {Hayes}}]{bib:HayesMag}%
  \BibitemOpen
  \bibfield  {author} {\bibinfo {author} {\bibfnamefont {X.}~\bibnamefont
  {Wang}}, \bibinfo {author} {\bibfnamefont {J.}~\bibnamefont {Friar}}, \ and\
  \bibinfo {author} {\bibfnamefont {A.}~\bibnamefont {Hayes}},\ }\href@noop {}
  {\bibfield  {journal} {\bibinfo  {journal} {Phys. Rev. C}\ ,\ \bibinfo
  {pages} {034314}} (\bibinfo {year} {2016})}\BibitemShut {NoStop}%
\bibitem [{\citenamefont {Mampe}\ \emph {et~al.}(1978)\citenamefont {Mampe},
  \citenamefont {Schreckenbach}, \citenamefont {Jeuch}, \citenamefont {Maier},
  \citenamefont {Braumandl}, \citenamefont {Larysz},\ and\ \citenamefont {von
  Egidy}}]{bib:BILL}%
  \BibitemOpen
  \bibfield  {author} {\bibinfo {author} {\bibfnamefont {W.}~\bibnamefont
  {Mampe}}, \bibinfo {author} {\bibfnamefont {K.}~\bibnamefont
  {Schreckenbach}}, \bibinfo {author} {\bibfnamefont {P.}~\bibnamefont
  {Jeuch}}, \bibinfo {author} {\bibfnamefont {B.}~\bibnamefont {Maier}},
  \bibinfo {author} {\bibfnamefont {F.}~\bibnamefont {Braumandl}}, \bibinfo
  {author} {\bibfnamefont {J.}~\bibnamefont {Larysz}}, \ and\ \bibinfo {author}
  {\bibfnamefont {T.}~\bibnamefont {von Egidy}},\ }\href@noop {} {\bibfield
  {journal} {\bibinfo  {journal} {Nucl. Inst. Meth.}\ }\textbf {\bibinfo
  {volume} {A154}},\ \bibinfo {pages} {127} (\bibinfo {year}
  {1978})}\BibitemShut {NoStop}%
\bibitem [{\citenamefont {Schreckenbach}\ \emph {et~al.}(1985)\citenamefont
  {Schreckenbach}, \citenamefont {Colvin}, \citenamefont {Gelletly},\ and\
  \citenamefont {Von~Feilitzsch}}]{bib:ILL_1}%
  \BibitemOpen
  \bibfield  {author} {\bibinfo {author} {\bibfnamefont {K.}~\bibnamefont
  {Schreckenbach}}, \bibinfo {author} {\bibfnamefont {G.}~\bibnamefont
  {Colvin}}, \bibinfo {author} {\bibfnamefont {W.}~\bibnamefont {Gelletly}}, \
  and\ \bibinfo {author} {\bibfnamefont {F.}~\bibnamefont {Von~Feilitzsch}},\
  }\href {\doibase 10.1016/0370-2693(85)91337-1} {\bibfield  {journal}
  {\bibinfo  {journal} {Phys.Lett.}\ }\textbf {\bibinfo {volume} {B160}},\
  \bibinfo {pages} {325} (\bibinfo {year} {1985})}\BibitemShut {NoStop}%
\bibitem [{\citenamefont {Von~Feilitzsch}\ \emph {et~al.}(1982)\citenamefont
  {Von~Feilitzsch}, \citenamefont {Hahn},\ and\ \citenamefont
  {Schreckenbach}}]{bib:ILL_2}%
  \BibitemOpen
  \bibfield  {author} {\bibinfo {author} {\bibfnamefont {F.}~\bibnamefont
  {Von~Feilitzsch}}, \bibinfo {author} {\bibfnamefont {A.~A.}\ \bibnamefont
  {Hahn}}, \ and\ \bibinfo {author} {\bibfnamefont {K.}~\bibnamefont
  {Schreckenbach}},\ }\href {\doibase 10.1016/0370-2693(82)90622-0} {\bibfield
  {journal} {\bibinfo  {journal} {Phys.Lett.}\ }\textbf {\bibinfo {volume}
  {B118}},\ \bibinfo {pages} {162} (\bibinfo {year} {1982})}\BibitemShut
  {NoStop}%
\bibitem [{\citenamefont {Hahn}\ \emph {et~al.}(1989)\citenamefont {Hahn} \emph
  {et~al.}}]{bib:ILL_3}%
  \BibitemOpen
  \bibfield  {author} {\bibinfo {author} {\bibfnamefont {A.~A.}\ \bibnamefont
  {Hahn}} \emph {et~al.},\ }\href {\doibase 10.1016/0370-2693(89)91598-0}
  {\bibfield  {journal} {\bibinfo  {journal} {Phys.Lett.}\ }\textbf {\bibinfo
  {volume} {B218}},\ \bibinfo {pages} {365} (\bibinfo {year}
  {1989})}\BibitemShut {NoStop}%
\bibitem [{\citenamefont {Onillon}(2018)}]{bib:ILL_calib}%
  \BibitemOpen
  \bibfield  {author} {\bibinfo {author} {\bibfnamefont {A.}~\bibnamefont
  {Onillon}},\ }\href
  {https://neutrinos.llnl.gov/content/assets/docs/workshops/2018/AAP2018-ILL-spectra-normalization-Onillon.pdf}
  {\enquote {\bibinfo {title} {{Updated Flux and Spectral Predictions relevant
  to the RAA}},}\ }\bibinfo {howpublished} {{Applied Antineutrino Physics
  Workshop (AAP) 2018}} (\bibinfo {year} {2018})\BibitemShut {NoStop}%
\bibitem [{\citenamefont {Giunti}\ \emph {et~al.}(2017)\citenamefont {Giunti},
  \citenamefont {Ji}, \citenamefont {Laveder}, \citenamefont {Li},\ and\
  \citenamefont {Littlejohn}}]{Giunti:2017yid}%
  \BibitemOpen
  \bibfield  {author} {\bibinfo {author} {\bibfnamefont {C.}~\bibnamefont
  {Giunti}}, \bibinfo {author} {\bibfnamefont {X.~P.}\ \bibnamefont {Ji}},
  \bibinfo {author} {\bibfnamefont {M.}~\bibnamefont {Laveder}}, \bibinfo
  {author} {\bibfnamefont {Y.~F.}\ \bibnamefont {Li}}, \ and\ \bibinfo {author}
  {\bibfnamefont {B.~R.}\ \bibnamefont {Littlejohn}},\ }\href {\doibase
  10.1007/JHEP10(2017)143} {\bibfield  {journal} {\bibinfo  {journal} {JHEP}\
  }\textbf {\bibinfo {volume} {10}},\ \bibinfo {pages} {143} (\bibinfo {year}
  {2017})},\ \Eprint {http://arxiv.org/abs/1708.01133} {arXiv:1708.01133
  [hep-ph]} \BibitemShut {NoStop}%
%%CITATION = ARXIV:1708.01133;%%
\bibitem [{\citenamefont {Gariazzo}\ \emph {et~al.}(2018)\citenamefont
  {Gariazzo}, \citenamefont {Giunti}, \citenamefont {Laveder},\ and\
  \citenamefont {Li}}]{Gariazzo:2018mwd}%
  \BibitemOpen
  \bibfield  {author} {\bibinfo {author} {\bibfnamefont {S.}~\bibnamefont
  {Gariazzo}}, \bibinfo {author} {\bibfnamefont {C.}~\bibnamefont {Giunti}},
  \bibinfo {author} {\bibfnamefont {M.}~\bibnamefont {Laveder}}, \ and\
  \bibinfo {author} {\bibfnamefont {Y.~F.}\ \bibnamefont {Li}},\ }\href
  {\doibase 10.1016/j.physletb.2018.04.057} {\bibfield  {journal} {\bibinfo
  {journal} {Phys. Lett. B}\ }\textbf {\bibinfo {volume} {782}},\ \bibinfo
  {pages} {13} (\bibinfo {year} {2018})},\ \Eprint
  {http://arxiv.org/abs/1801.06467} {arXiv:1801.06467 [hep-ph]} \BibitemShut
  {NoStop}%
%%CITATION = ARXIV:1801.06467;%%
\bibitem [{\citenamefont {Dentler}\ \emph {et~al.}(2017)\citenamefont
  {Dentler}, \citenamefont {Hern{\'a}ndez-Cabezudo}, \citenamefont {Kopp},
  \citenamefont {Maltoni},\ and\ \citenamefont {Schwetz}}]{schwetz_global}%
  \BibitemOpen
  \bibfield  {author} {\bibinfo {author} {\bibfnamefont {M.}~\bibnamefont
  {Dentler}}, \bibinfo {author} {\bibfnamefont {A.}~\bibnamefont
  {Hern{\'a}ndez-Cabezudo}}, \bibinfo {author} {\bibfnamefont {J.}~\bibnamefont
  {Kopp}}, \bibinfo {author} {\bibfnamefont {M.}~\bibnamefont {Maltoni}}, \
  and\ \bibinfo {author} {\bibfnamefont {T.}~\bibnamefont {Schwetz}},\
  }\href@noop {} {\bibfield  {journal} {\bibinfo  {journal} {JHEP}\ }\textbf
  {\bibinfo {volume} {1711:099}} (\bibinfo {year} {2017})}\BibitemShut
  {NoStop}%
\bibitem [{\citenamefont {Giunti}(2017{\natexlab{b}})}]{Giunti:2017nww}%
  \BibitemOpen
  \bibfield  {author} {\bibinfo {author} {\bibfnamefont {C.}~\bibnamefont
  {Giunti}},\ }\href {\doibase 10.1103/PhysRevD.96.033005} {\bibfield
  {journal} {\bibinfo  {journal} {Phys. Rev.}\ }\textbf {\bibinfo {volume}
  {D96}},\ \bibinfo {pages} {033005} (\bibinfo {year} {2017}{\natexlab{b}})},\
  \Eprint {http://arxiv.org/abs/1704.02276} {arXiv:1704.02276 [hep-ph]}
  \BibitemShut {NoStop}%
%%CITATION = ARXIV:1704.02276;%%
\bibitem [{\citenamefont {An}\ \emph {et~al.}(2017)\citenamefont {An} \emph
  {et~al.}}]{bib:prl_evol}%
  \BibitemOpen
  \bibfield  {author} {\bibinfo {author} {\bibfnamefont {F.~P.}\ \bibnamefont
  {An}} \emph {et~al.} (\bibinfo {collaboration} {Daya Bay}),\ }\href@noop {}
  {\bibfield  {journal} {\bibinfo  {journal} {Phys. Rev. Lett.}\ }\textbf
  {\bibinfo {volume} {118}},\ \bibinfo {pages} {251801} (\bibinfo {year}
  {2017})}\BibitemShut {NoStop}%
\bibitem [{\citenamefont {Ko}\ \emph {et~al.}(2017)\citenamefont {Ko} \emph
  {et~al.}}]{bib:neos}%
  \BibitemOpen
  \bibfield  {author} {\bibinfo {author} {\bibfnamefont {Y.~J.}\ \bibnamefont
  {Ko}} \emph {et~al.},\ }\href@noop {} {\bibfield  {journal} {\bibinfo
  {journal} {Phys. Rev. Lett.}\ }\textbf {\bibinfo {volume} {118}},\ \bibinfo
  {pages} {121802} (\bibinfo {year} {2017})}\BibitemShut {NoStop}%
\bibitem [{\citenamefont {Alekseev}\ \emph {et~al.}(2018)\citenamefont
  {Alekseev} \emph {et~al.}}]{danss_osc}%
  \BibitemOpen
  \bibfield  {author} {\bibinfo {author} {\bibfnamefont {I.}~\bibnamefont
  {Alekseev}} \emph {et~al.},\ }\href@noop {} {\  (\bibinfo {year} {2018})},\
  \Eprint {http://arxiv.org/abs/1804.04046} {arXiv:1804.04046 [hep-ex]}
  \BibitemShut {NoStop}%
\bibitem [{\citenamefont {Ashenfelter}\ \emph {et~al.}(2018)\citenamefont
  {Ashenfelter} \emph {et~al.}}]{prospect_osc}%
  \BibitemOpen
  \bibfield  {author} {\bibinfo {author} {\bibfnamefont {J.}~\bibnamefont
  {Ashenfelter}} \emph {et~al.} (\bibinfo {collaboration} {PROSPECT}),\
  }\href@noop {} {\  (\bibinfo {year} {2018})},\ \Eprint
  {http://arxiv.org/abs/1806.02784} {arXiv:1806.02784 [hep-ex]} \BibitemShut
  {NoStop}%
\bibitem [{\citenamefont {Lhuillier}(2015)}]{stereo}%
  \BibitemOpen
  \bibfield  {author} {\bibinfo {author} {\bibfnamefont {D.}~\bibnamefont
  {Lhuillier}},\ }\href@noop {} {\bibfield  {journal} {\bibinfo  {journal} {AIP
  Conf. Proc.}\ }\textbf {\bibinfo {volume} {1666}},\ \bibinfo {pages} {180003}
  (\bibinfo {year} {2015})}\BibitemShut {NoStop}%
\bibitem [{\citenamefont {Bak}\ \emph {et~al.}(2018)\citenamefont {Bak} \emph
  {et~al.}}]{RenoEvol}%
  \BibitemOpen
  \bibfield  {author} {\bibinfo {author} {\bibfnamefont {G.}~\bibnamefont
  {Bak}} \emph {et~al.},\ }\href@noop {} {\  (\bibinfo {year} {2018})},\
  \Eprint {http://arxiv.org/abs/1806.00574} {arXiv:1806.00574 [hep-ex]}
  \BibitemShut {NoStop}%
\bibitem [{\citenamefont {An}\ \emph {et~al.}(2018)\citenamefont {An} \emph
  {et~al.}}]{bib:prd_flux}%
  \BibitemOpen
  \bibfield  {author} {\bibinfo {author} {\bibfnamefont {F.~P.}\ \bibnamefont
  {An}} \emph {et~al.} (\bibinfo {collaboration} {Daya Bay}),\ }\href@noop {}
  {\  (\bibinfo {year} {2018})},\ \Eprint {http://arxiv.org/abs/1808.10836}
  {arXiv:1808.10836 [hep-ex]} \BibitemShut {NoStop}%
\bibitem [{\citenamefont {Bezerra}(2018)}]{bib:dc_flux}%
  \BibitemOpen
  \bibfield  {author} {\bibinfo {author} {\bibfnamefont {T.}~\bibnamefont
  {Bezerra}},\ }\href {http://www.ba.infn.it/~now/now2018/program.html}
  {\enquote {\bibinfo {title} {{Double Chooz Results at the Double Detector
  Phase}},}\ }\bibinfo {howpublished} {{Neutrino Oscillation Workshop (NOW)
  2018}} (\bibinfo {year} {2018})\BibitemShut {NoStop}%
\bibitem [{\citenamefont {Kwon}\ \emph {et~al.}(1981)\citenamefont {Kwon} \emph
  {et~al.}}]{bib:ILL_nu}%
  \BibitemOpen
  \bibfield  {author} {\bibinfo {author} {\bibfnamefont {H.}~\bibnamefont
  {Kwon}} \emph {et~al.},\ }\href@noop {} {\bibfield  {journal} {\bibinfo
  {journal} {Phys. Rev. D}\ }\textbf {\bibinfo {volume} {24}},\ \bibinfo
  {pages} {1097} (\bibinfo {year} {1981})}\BibitemShut {NoStop}%
\bibitem [{\citenamefont {Hoummada}\ \emph {et~al.}(1995)\citenamefont
  {Hoummada}, \citenamefont {Mikou}, \citenamefont {Avenier}, \citenamefont
  {Bagieu}, \citenamefont {Cavaignac},\ and\ \citenamefont
  {Koang}}]{ILL_nuFix}%
  \BibitemOpen
  \bibfield  {author} {\bibinfo {author} {\bibfnamefont {A.}~\bibnamefont
  {Hoummada}}, \bibinfo {author} {\bibfnamefont {S.~L.}\ \bibnamefont {Mikou}},
  \bibinfo {author} {\bibfnamefont {M.}~\bibnamefont {Avenier}}, \bibinfo
  {author} {\bibfnamefont {G.}~\bibnamefont {Bagieu}}, \bibinfo {author}
  {\bibfnamefont {J.}~\bibnamefont {Cavaignac}}, \ and\ \bibinfo {author}
  {\bibfnamefont {D.~H.}\ \bibnamefont {Koang}},\ }\href@noop {} {\bibfield
  {journal} {\bibinfo  {journal} {Applied Radiation and Isotopes}\ }\textbf
  {\bibinfo {volume} {46}},\ \bibinfo {pages} {449} (\bibinfo {year}
  {1995})}\BibitemShut {NoStop}%
\bibitem [{\citenamefont {Greenwood}\ \emph {et~al.}(1996)\citenamefont
  {Greenwood} \emph {et~al.}}]{bib:srp}%
  \BibitemOpen
  \bibfield  {author} {\bibinfo {author} {\bibfnamefont {Z.}~\bibnamefont
  {Greenwood}} \emph {et~al.},\ }\href@noop {} {\bibfield  {journal} {\bibinfo
  {journal} {Phys. Rev. D}\ }\textbf {\bibinfo {volume} {53}},\ \bibinfo
  {pages} {6054} (\bibinfo {year} {1996})}\BibitemShut {NoStop}%
\bibitem [{\citenamefont {Vidyakin}\ \emph {et~al.}(1987)\citenamefont
  {Vidyakin} \emph {et~al.}}]{bib:Krasno1}%
  \BibitemOpen
  \bibfield  {author} {\bibinfo {author} {\bibfnamefont {G.~S.}\ \bibnamefont
  {Vidyakin}} \emph {et~al.},\ }\href@noop {} {\bibfield  {journal} {\bibinfo
  {journal} {Sov. Phys. JETP}\ }\textbf {\bibinfo {volume} {66}},\ \bibinfo
  {pages} {243} (\bibinfo {year} {1987})}\BibitemShut {NoStop}%
\bibitem [{\citenamefont {Vidyakin}\ \emph {et~al.}(1990)\citenamefont
  {Vidyakin} \emph {et~al.}}]{bib:Krasno2}%
  \BibitemOpen
  \bibfield  {author} {\bibinfo {author} {\bibfnamefont {G.~S.}\ \bibnamefont
  {Vidyakin}} \emph {et~al.},\ }\href@noop {} {\bibfield  {journal} {\bibinfo
  {journal} {Sov. Phys. JETP}\ }\textbf {\bibinfo {volume} {71}},\ \bibinfo
  {pages} {424} (\bibinfo {year} {1990})}\BibitemShut {NoStop}%
\bibitem [{\citenamefont {Vidyakin}\ \emph {et~al.}(1994)\citenamefont
  {Vidyakin} \emph {et~al.}}]{bib:Krasno4}%
  \BibitemOpen
  \bibfield  {author} {\bibinfo {author} {\bibfnamefont {G.~S.}\ \bibnamefont
  {Vidyakin}} \emph {et~al.},\ }\href@noop {} {\bibfield  {journal} {\bibinfo
  {journal} {JETP Lett}\ }\textbf {\bibinfo {volume} {59}},\ \bibinfo {pages}
  {390} (\bibinfo {year} {1994})}\BibitemShut {NoStop}%
\bibitem [{\citenamefont {Boireau}\ \emph {et~al.}(2016)\citenamefont {Boireau}
  \emph {et~al.}}]{bib:nucifer}%
  \BibitemOpen
  \bibfield  {author} {\bibinfo {author} {\bibfnamefont {G.}~\bibnamefont
  {Boireau}} \emph {et~al.} (\bibinfo {collaboration} {NUCIFER}),\ }\href@noop
  {} {\bibfield  {journal} {\bibinfo  {journal} {Phys. Rev. D}\ }\textbf
  {\bibinfo {volume} {93}},\ \bibinfo {pages} {112006} (\bibinfo {year}
  {2016})}\BibitemShut {NoStop}%
\bibitem [{\citenamefont {Zacek}\ \emph {et~al.}(1986)\citenamefont {Zacek}
  \emph {et~al.}}]{bib:gosgen}%
  \BibitemOpen
  \bibfield  {author} {\bibinfo {author} {\bibfnamefont {G.}~\bibnamefont
  {Zacek}} \emph {et~al.} (\bibinfo {collaboration} {CalTech-SIN-TUM}),\
  }\href@noop {} {\bibfield  {journal} {\bibinfo  {journal} {Phys. Rev. D}\
  }\textbf {\bibinfo {volume} {34}},\ \bibinfo {pages} {2621} (\bibinfo {year}
  {1986})}\BibitemShut {NoStop}%
\bibitem [{\citenamefont {Afonin}\ \emph {et~al.}(1988)\citenamefont {Afonin}
  \emph {et~al.}}]{bib:Rovno1}%
  \BibitemOpen
  \bibfield  {author} {\bibinfo {author} {\bibfnamefont {A.}~\bibnamefont
  {Afonin}} \emph {et~al.},\ }\href@noop {} {\bibfield  {journal} {\bibinfo
  {journal} {Sov. Phys. JETP}\ }\textbf {\bibinfo {volume} {67}},\ \bibinfo
  {pages} {213} (\bibinfo {year} {1988})}\BibitemShut {NoStop}%
\bibitem [{\citenamefont {Kuvshinnikov}\ \emph {et~al.}(1991)\citenamefont
  {Kuvshinnikov} \emph {et~al.}}]{bib:Rovno2}%
  \BibitemOpen
  \bibfield  {author} {\bibinfo {author} {\bibfnamefont {A.}~\bibnamefont
  {Kuvshinnikov}} \emph {et~al.},\ }\href@noop {} {\bibfield  {journal}
  {\bibinfo  {journal} {JETP Lett.}\ }\textbf {\bibinfo {volume} {54}},\
  \bibinfo {pages} {253} (\bibinfo {year} {1991})}\BibitemShut {NoStop}%
\bibitem [{\citenamefont {Achkar}\ \emph {et~al.}(1995)\citenamefont {Achkar}
  \emph {et~al.}}]{bib:Bugey3_osc}%
  \BibitemOpen
  \bibfield  {author} {\bibinfo {author} {\bibfnamefont {B.}~\bibnamefont
  {Achkar}} \emph {et~al.},\ }\href@noop {} {\bibfield  {journal} {\bibinfo
  {journal} {Nucl. Phys. B}\ }\textbf {\bibinfo {volume} {434}},\ \bibinfo
  {pages} {503} (\bibinfo {year} {1995})}\BibitemShut {NoStop}%
\bibitem [{\citenamefont {Declais}\ \emph {et~al.}(1994)\citenamefont {Declais}
  \emph {et~al.}}]{bib:B4}%
  \BibitemOpen
  \bibfield  {author} {\bibinfo {author} {\bibfnamefont {Y.}~\bibnamefont
  {Declais}} \emph {et~al.},\ }\href@noop {} {\bibfield  {journal} {\bibinfo
  {journal} {Phys. Lett. B}\ }\textbf {\bibinfo {volume} {338}},\ \bibinfo
  {pages} {383} (\bibinfo {year} {1994})}\BibitemShut {NoStop}%
\bibitem [{\citenamefont {Boehm}\ \emph {et~al.}(2001)\citenamefont {Boehm}
  \emph {et~al.}}]{bib:paloverde}%
  \BibitemOpen
  \bibfield  {author} {\bibinfo {author} {\bibfnamefont {F.}~\bibnamefont
  {Boehm}} \emph {et~al.},\ }\href@noop {} {\bibfield  {journal} {\bibinfo
  {journal} {Phys. Rev. D}\ }\textbf {\bibinfo {volume} {64}},\ \bibinfo
  {pages} {112001} (\bibinfo {year} {2001})}\BibitemShut {NoStop}%
\bibitem [{\citenamefont {Apollonio}\ \emph {et~al.}(2003)\citenamefont
  {Apollonio} \emph {et~al.}}]{bib:chooz2}%
  \BibitemOpen
  \bibfield  {author} {\bibinfo {author} {\bibfnamefont {M.}~\bibnamefont
  {Apollonio}} \emph {et~al.},\ }\href@noop {} {\bibfield  {journal} {\bibinfo
  {journal} {Eur. J. Phys C}\ }\textbf {\bibinfo {volume} {27}},\ \bibinfo
  {pages} {331} (\bibinfo {year} {2003})}\BibitemShut {NoStop}%
\bibitem [{\citenamefont {Gariazzo}\ \emph {et~al.}(2017)\citenamefont
  {Gariazzo}, \citenamefont {Giunti}, \citenamefont {Laveder},\ and\
  \citenamefont {Li}}]{Gariazzo:2017fdh}%
  \BibitemOpen
  \bibfield  {author} {\bibinfo {author} {\bibfnamefont {S.}~\bibnamefont
  {Gariazzo}}, \bibinfo {author} {\bibfnamefont {C.}~\bibnamefont {Giunti}},
  \bibinfo {author} {\bibfnamefont {M.}~\bibnamefont {Laveder}}, \ and\
  \bibinfo {author} {\bibfnamefont {Y.~F.}\ \bibnamefont {Li}},\ }\href
  {\doibase 10.1007/JHEP06(2017)135} {\bibfield  {journal} {\bibinfo  {journal}
  {JHEP}\ }\textbf {\bibinfo {volume} {06}},\ \bibinfo {pages} {135} (\bibinfo
  {year} {2017})},\ \Eprint {http://arxiv.org/abs/1703.00860} {arXiv:1703.00860
  [hep-ph]} \BibitemShut {NoStop}%
%%CITATION = ARXIV:1703.00860;%%
\end{thebibliography}%

\end{document}